\documentclass[twocolappendix,numberedappendix]{emulateapj} 
\usepackage{tabularx, booktabs}
\usepackage{natbib,amsmath,booktabs,apjfonts,threeparttable,multirow,graphicx,rotating}
\usepackage{adjustbox}
\usepackage[colorlinks=true,linkcolor=blue,citecolor=blue,urlcolor=blue]{hyperref}
\slugcomment{Accepted to ApJ}
\newcommand{\ltsima}{$\; \buildrel < \over \sim \;$}
\newcommand{\simlt}{\lower.5ex\hbox{\ltsima}}

\newcommand{\ls}{{_<\atop^{\sim}}}

\newcommand{\gs}{{_>\atop^{\sim}}}

\def\arcdeg{\hbox{$^\circ$}}
\def\arcmin{\hbox{$^\prime$}}
\def\arcsec{\hbox{$^{\prime\prime}$}}

\begin{document}
	
\title{Measurement of the core-collapse progenitor mass distribution of the Small Magellanic Cloud}
	
\author{Katie Auchettl\altaffilmark{1,2,6}, Laura A. Lopez\altaffilmark{1,3,6}, Carles Badenes\altaffilmark{4}, Enrico Ramirez-Ruiz\altaffilmark{5,6}, John F. Beacom\altaffilmark{1,2,3}and Tyler Holland-Ashford\altaffilmark{1,3}}
	
\altaffiltext{1}{Center for Cosmology and AstroParticle Physics (CCAPP), The Ohio State University, 191 W. Woodruff Ave., Columbus, OH 43210, USA}
\altaffiltext{2}{Department of Physics, The Ohio State University, 191 W. Woodruff Ave., Columbus, OH 43210, USA}
\altaffiltext{3}{Department of Astronomy, The Ohio State University, 140 West 18th Avenue, Columbus, OH 43210, USA}
\altaffiltext{4}{Department of Physics and Astronomy, University of Pittsburgh, 3941 O'Hara Street, Pittsburgh, PA 15260, USA}
\altaffiltext{5}{Department of Astronomy and Astrophysics, University of California, Santa Cruz, CA 95064, USA}
\altaffiltext{6}{Dark Cosmology Centre, Niels Bohr Institute, University of Copenhagen, Blegdamsvej 17, 2100 Copenhagen, Denmark}

\begin{abstract}
The physics of core-collapse (CC) supernovae (SNe) and how the explosions depend on progenitor properties are central questions in astronomy. For only a handful of SNe, the progenitor star has been identified in pre-explosion images. Supernova remnants (SNRs), which are observed long after the original SN event, provide a unique opportunity to increase the number of progenitor measurements. Here, we systematically examine the stellar populations in the vicinities of 23 known SNRs in the Small Magellanic Cloud (SMC) using the star formation history (SFH) maps of \citet{2004AJ....127.1531H}. We combine the results with constraints on the SNR metal abundances and environment from X-ray and optical observations. We find that 22 SNRs in the SMC have local SFHs and properties consistent with a CC explosion, several of which are likely to have been high-mass progenitors. This result supports recent theoretical findings that high-mass progenitors can produce successful explosions. We estimate the mass distribution of the CC progenitors and find that this distribution is similar to a Salpeter IMF (within the uncertainties), while this result is shallower than the mass distribution found in M31 and M33 by \cite{2014ApJ...795..170J} and \cite{2018arXiv180207870D} using a similar approach. Additionally, we find that a number of the SMC SNRs exhibit a burst of star formation between 50--200 Myr ago. As these sources are likely CC, this signature may be indicative of massive stars undergoing delayed CC as a consequence of binary interaction, rapid rotation, or low metallicity. In addition, the lack of Type Ia SNRs in the SMC is possibly a result of the short visibility times of these sources as they may fall below the sensitivity limits of current radio observations. 
\end{abstract}
	
	
\section{Introduction}

One of the most uncertain aspects of stellar evolution is the link between the nature of supernovae (SNe) and their progenitor stars. Generally, it is thought that massive stars ($\gtrsim8M_{\odot}$) undergo core collapse (CC) \citep[e.g.,][]{1974ARA&A..12..215I, 2002RvMP...74.1015W, 2004MNRAS.353...87E, 2009ARA&A..47...63S, 2012ApJ...761...26J, 2013ApJ...765L..43I, 2018arXiv180207870D} once nuclear burning has led to an iron core, and their explosions are driven by a combination of neutrino heating, turbulence, and convection (e.g., reviews by e.g., \citealt{2016PASA...33...48M} and \citealt{2017arXiv170208825J}, as well as \citealt{2013ApJ...771...52M, 2015ApJ...799....5C, 2015ApJ...800...10D, 2015A&A...577A..48W}). However, the type of SN explosion \citep[i.e.,  Type Ib, Ic, and II; see review by e.g.,][]{2016arXiv161109353G} is expected to depend strongly on the mass of the progenitor \citep[e.g.,][]{2002RvMP...74.1015W, 2003ApJ...591..288H}, whether it is found in a binary \citep{1998A&A...333..557D, 2004MNRAS.348.1215I, 2017arXiv170107032Z}, factors such as the metallicity and rotation of the progenitor star \citep[e.g.,][]{2003ApJ...591..288H, 2004MNRAS.353...87E, 2005ApJ...620..861T, 2012ApJ...749...91F, 2017MNRAS.467.3347K}, and a combination of all these effects \citep[][]{2005A&A...443..581H,2007A&A...461..571H, 2008MNRAS.384.1109E,2017PASA...34...58E}. As such, it is difficult for current theory to confidently predict which stars undergo CC and produce a SN of a given type and which of those produce a neutron star or a black hole \citep[e.g.,][]{2012ApJ...757...69U, 2014ApJ...783...10S, 2016ApJ...821...38S,2018arXiv180403182E}.

There have been a number of attempts to identify the progenitor stars of CC SNe in preceding space- and ground-based images \citep[see reviews by][and references therein]{2009ARA&A..47...63S, 2015PASA...32...16S}. However, even though direct imaging addresses the above challenges, this method is limited by a number of factors, including the SN rate in the local universe and the depth of archival and/or current observations. Consequently, only a handful of SNe have detections of their progenitor stars (see e.g., review by \citealt{2015PASA...32...16S} and studies such as \citealt{2015MNRAS.450.3289G, 2015MNRAS.452.2195A, 2017ApJ...836..222F, 2018ApJ...860...90V}).

Supernova remnants (SNRs), structures resulting from SNe that occurred hundreds or thousands of years ago, provide us with a unique opportunity to greatly increase the number of explosions with progenitor constraints. The most common method used to link SNRs to their explosions is to estimate metal abundances based on X-ray emission line strengths and to compare these values to those predicted in SN models of different mass progenitors \citep[see e.g.,][]{2006ApJ...645.1373B, 2015ApJ...801L..31Y, 2015ApJ...803..101P}. In rare cases, light echoes from the originating SNe associated with young SNRs can be observed, providing more information about the progenitor \citep[e.g.,][]{2005Natur.438.1132R, 2008ApJ...680.1137R, 2008ApJ...681L..81R}. 

Another approach is to characterise progenitor properties of SNRs by examining the resolved stellar populations in their proximity \citep[e.g.,][]{2009ApJ...700..727B, 2012ApJ...761...26J, 2014ApJ...795..170J,2018arXiv180207870D}. These works exploit the campaigns to probe the star-formation histories (SFHs) across nearby galaxies, such as the Large Magellanic Cloud (LMC), M31, and M33 \citep[e.g.,][]{2004AJ....127.1531H, 2009AJ....138.1243H, 2015ApJ...805..183L}. This approach provides a nearly complete (up to $\sim$70\%) view of the SN progenitors that exploded in these galaxies \citep[see detailed study by][]{2017MNRAS.464.2326S}, which is important for characterising SN rates in the Local Group and how they vary in different galactic environments (e.g., with metallicity). 

Combining these two methods, \citet{2009ApJ...700..727B} used the SFH maps by \citet{2009AJ....138.1243H} of the LMC, to estimate the progenitor ages and masses of four CC SNRs and four Type Ia SNRs. They found that their sample of CC SNRs are associated with vigorous star formation within the last few Myr. However, Type Ia SNRs, which result from the thermonuclear explosion of a white dwarf (WD) in a binary system with either a non-degenerate companion star (single degenerate scenario) or another WD (double degenerate scenario) \citep[see review by e.g., ][]{2014ARA&A..52..107M}, are located in a wide range of environments,  including both old, metal-poor populations and recent, highly star-forming regions. Generally, the presence of (or lack of) recent star formation around a given SNR tentatively suggests a CC (Type Ia) origin, although the averaged SFHs derived from resolved stellar populations in SNR vicinities can be misleading, especially for Type Ia SNRs. However, when combined with constraints on the SNR metal abundances and environment, the local SFHs at the sites of SNRs are a useful, complementary tool to gain insights about the SN progenitors.

In this paper, we examine the SFH maps of the SMC derived by \citet{2004AJ....127.1531H} at the sites of the 23 known SNRs in that galaxy \citep{2010MNRAS.407.1301B}. Combined with constraints on the metal abundances and environment of each SNR derived from optical and X-ray observations, we attempt to connect the SNR to its progenitor. The paper is organised as follows. In Section \ref{sect1}, we summarise the properties of the SMC and review its known SNRs. In Section \ref{sect2}, we discuss the SFH map of the SMC derived by \citet{2004AJ....127.1531H} and the galaxy's SFH as a whole. In Section \ref{sect3}, we outline the selection criteria we used to choose our sample. In Section \ref{sect4}, we examine the SFH around each SNR in our sample and compare them to the properties of the SNe derived from the SNRs themselves. In Section \ref{sect5}, we discuss these results in the context of our understanding of the progenitors of SNe, and in Section \ref{sect6} we present our conclusions. We include an appendix summarising the observational properties of each SNR as a resource for the reader. 

\section{The Small Magellanic Cloud and its supernova remnants}\label{sect1}

\begin{figure}[t]
	\includegraphics[width=\columnwidth]{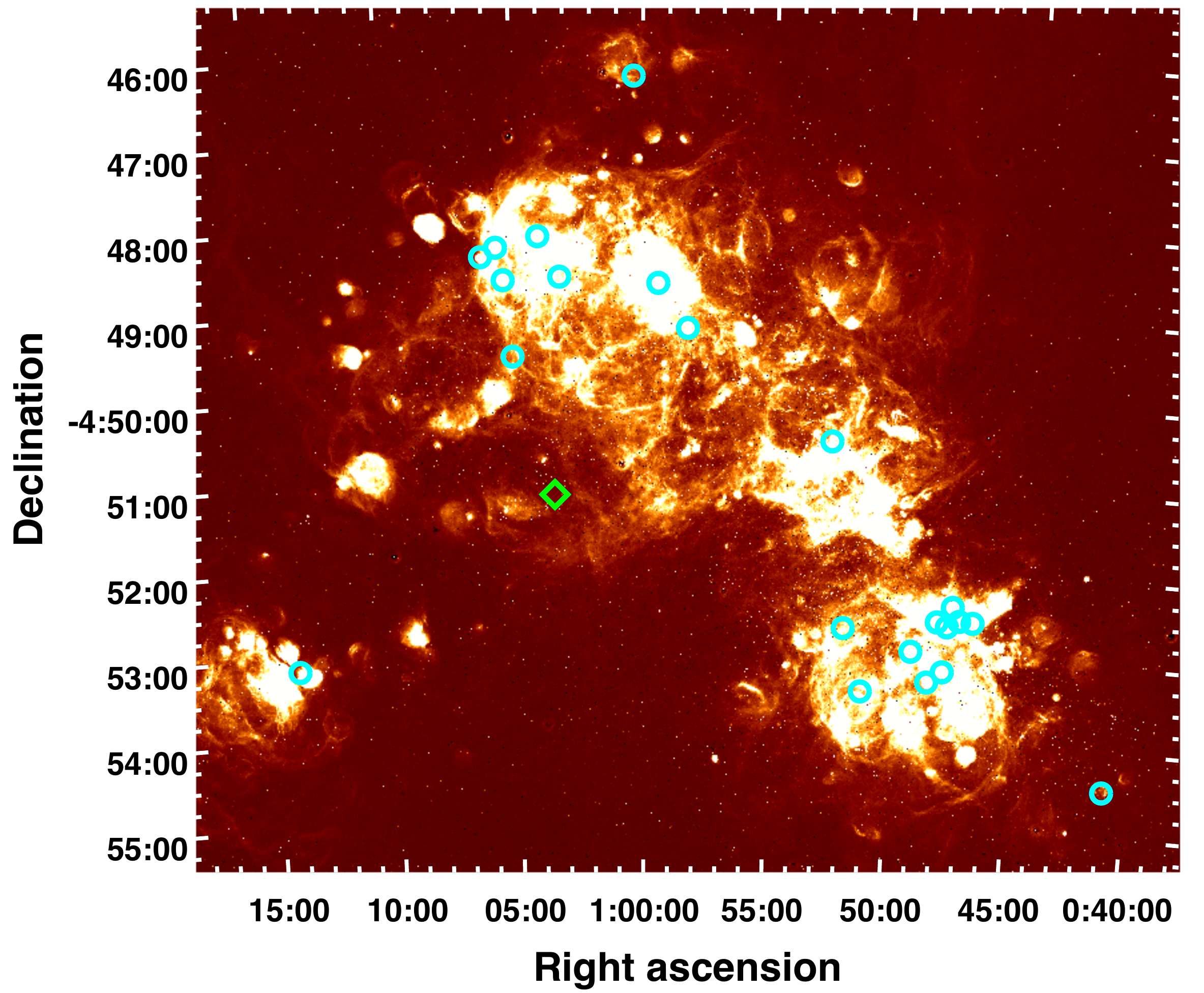}
	\caption{Continuum-subtracted H$\alpha$ image of the SMC from MCELS \citep{smith99,winkler15}. Here the cyan circular regions marking the locations of the 22 SMC SNRs that based on our analysis have SFHs and properties consistent with that of a CC progenitor. The green diamond marks the position of HFPK~334 whose SFH and properties seem to imply a possible Type Ia progenitor (see Section \ref{sect4} for more details). \label{SMC}}
\end{figure}

The SMC is a close ($\sim$61~kpc away: \citealt{hilditch05}), metal-poor galaxy (with $Z_{\rm SMC} \sim Z_{\sun}/5$: \citealt{dufour84,dopita92}) with active star formation \citep{wilke04,bolatto11,skibba12}. The SMC has a diverse population of 23 SNRs that have been identified at several wavelengths \citep{2010MNRAS.407.1301B,2015ApJ...799..158T}, and extensive work has gone into characterising the properties and nature of these sources (e.g., \citealt{2004A&A...421.1031V,haberl12,2014ApJ...788....5L,2016PASJ...68S...9T}). In Figure \ref{SMC}, we plot as a green diamond and cyan circles the positions of all 23 SNRs on a continuum-subtracted H$\alpha$ image of the SMC from the Magellanic Cloud Emission Line Survey \citep{smith99,winkler15}. In Table~\ref{listofSNRs}, we list the sources' names, positions, and suggested explosive origin of each source. In the appendix, we collate and summarise the properties of each remnant as presented in the literature, providing an up-to date review of all known SNRs in the SMC. These characteristics are used later in this paper to verify the classification of these sources as either Type Ia or CC based on their SFHs.

Of these remnants, 4 have possible evidence for Type Ia explosive origins presented in the literature (IKT~4 [SNR J0048.4$-$7319]; IKT~5 [SNR 0047$-$73.5]; IKT~25 [SNR 0104$-$72.3]; DEM~S128 [SNR 0103$-$72.4]), based on strong Fe-L emission in their X-ray spectra \citep{2004A&A...421.1031V}. However, others have argued that their environments and intermediate-mass element abundances are more consistent with CC SNR models (e.g., \citealt{2014ApJ...788....5L}). Out of the 23 SMC SNRs, 8 SNRs have evidence for CC origin (e.g., enhanced intermediate-mass element abundances, detection of a pulsar): DEM~S32 [SNR J0046.6$-$7309], IKT~2 [B0045-73.4], HFPK~419 [SNR J0047.7$-$7310], IKT~6 [B0049$-$73.6], IKT~7 [SNR J0051.9$-$7310], IKT~16 [SNR J0058.3$-$7218], B0102$-$7219 [SNR 0102$-$72.3], and IKT~23 [SNR 0103$-$72.6]. The other 11 SNRs identified in the SMC do not have explosion classifications (as Type Ia or CC) in the literature.

Out of the 23 SNRs in the SMC, IKT~25 is the only source where the SFH has been used to constrain the progenitor. \citet{2014ApJ...788....5L} found that the peak star formation rate at the site of IKT~25 is consistent with a progenitor of a mass $\sim$20--40$M_{\sun}$. This result was consistent with the intermediate-mass element abundances and the morphological characteristics found using deep {\it Chandra} X-ray Observatory observations. 

\begin{table*}[]
\centering
\caption{SMC supernova remnants and their literature suggested origins. Here, we have listed the SNRs in order of right ascension. Table adapted from \citet{2010MNRAS.407.1301B} and \citet{2015ApJ...799..158T}.}
\label{listofSNRs}
\resizebox{\textwidth}{!}{
\begin{tabular}{lp{6cm}cccccccc}
\toprule
Name & Alternative Name  & R.A. (J2000) & Dec. (J2000) & Diameter & Suggested &Reference\\ 
 & & & &(arcsec)& SNe origin & \\ \midrule
DEM~S5	&	B0039--7353, HFPK~530, SNR~J0040.9--7337	&	$	00	^{h}	40	^{m}	55	^{s}	$	&	$	-73	^{d}	36	^{m}	55	^{s}	$	&	121	 & 	? &	\\
DEM~S32	&	SNR~J0046.6--7309	&	$	00	^{h}	46	^{m}	39	^{s}	$	&	$	-73	^{d}	08	^{m}	39	^{s}	$	&	136	 & 	CC & [1]	\\
IKT~2	&	SNR~J0047.2--7308, HFPK~413, B0045--73.4	&	$	00	^{h}	47	^{m}	12	^{s}	$	&	$	-73	^{d}	08	^{m}	26	^{s}	$	&	66	 & 	CC & [1]	\\
B0045--733	&	HFPK~401, SNR~J0047.5--7306	&	$	00	^{h}	47	^{m}	29	^{s}	$	&	$	-73	^{d}	06	^{m}	01	^{s}	$	&	180	 & 	?&	\\
HFPK~419	&	SNR~J0047.7--7310	&	$	00	^{h}	47	^{m}	41	^{s}	$	&	$	-73	^{d}	09	^{m}	30	^{s}	$	&	90	 & 	CC &[1]	\\
NS~21	&	DEM~S35, SNR~J0047.8--7317	&	$	00	^{h}	47	^{m}	48	^{s}	$	&	$	-73	^{d}	17	^{m}	27	^{s}	$	&	76	 & 	?&	\\
NS~19	&	DEM~S31, SNR~J0048.1--7309	&	$	00	^{h}	48	^{m}	06	^{s}	$	&	$	-73	^{d}	08	^{m}	43	^{s}	$	&	79	 & 	?&	\\
IKT~4	&	HFPK~454, SNR~J0048.4--7319	&	$	00	^{h}	48	^{m}	25	^{s}	$	&	$	-73	^{d}	19	^{m}	24	^{s}	$	&	84	 & 	Ia &[1]	\\
IKT~5	&	DEM~S49, HFPK~437, SNR~0047--73.5, SNR~J0049.1--7314	&	$	00	^{h}	49	^{m}	07	^{s}	$	&	$	-73	^{d}	14	^{m}	05	^{s}	$	&	116	 & 	Ia/CC &[1, 2, 3]	\\
IKT~6	&	1E~0049.4--7339, HFPK~461, B0049--73.6, SNR~J0051.1--7321	&	$	00	^{h}	51	^{m}	07	^{s}	$	&	$	-73	^{d}	21	^{m}	26	^{s}	$	&	144	 & 	CC& [1,4, 5]	\\
IKT~7	&	HFPK~424, SNR~J0051.9--7310	&	$	00	^{h}	51	^{m}	54	^{s}	$	&	$	-73	^{d}	10	^{m}	24	^{s}	$	&	97	 & 	CC &[6]	\\
B0050--728	&	DEM~S68SE, HFPK~285, SNR~J0052.6--7238, SNR~J005240--723820, SMC~258, NS76	&	$	00	^{h}	52	^{m}	33	^{s}	$	&	$	-72	^{d}	37	^{m}	35	^{s}	$	&	323	 & 	?&	\\
IKT~16	&	HFPK~185/194, SNR~J0058.3--7218	&	$	00	^{h}	58	^{m}	16	^{s}	$	&	$	-72	^{d}	18	^{m}	05	^{s}	$	&	200	 & 	CC & [7,8]	\\
IKT~18	&	1E~0057.6--7228, HFPK~148, SNR~J0059.4--7210	&	$	00	^{h}	59	^{m}	25	^{s}	$	&	$	-72	^{d}	10	^{m}	10	^{s}	$	&	158	 & 	CC &[9]	\\
DEM~S108	&	B0058--71.8, HFPK~45, SNR~J0100.3--7134	&	$	01	^{h}	00	^{m}	21	^{s}	$	&	$	-71	^{d}	33	^{m}	40	^{s}	$	&	149	 & 	?&	\\
IKT~21	&	1E~0101.5--7226, HFPK~143, B0101--72.4, SNR~J0103.2--7209	&	$	01	^{h}	03	^{m}	13	^{s}	$	&	$	-72	^{d}	08	^{m}	59	^{s}	$	&	62	 & 	CC& [6]	\\
HFPK~334	&	SNR~J0103.5--7247	&	$	01	^{h}	03	^{m}	30	^{s}	$	&	$	-72	^{d}	47	^{m}	20	^{s}	$	&	86	 & 	? &	\\
1E~0102.2--7219&	DEM~S124, IKT~22, 1E0102--72.3, HFPK~107, SNR~0102--72.3, SNR~J0104.0--7202, B0102--7219&	$	01	^{h}	04	^{m}	02	^{s}	$	&	$	-72	^{d}	01	^{m}	48	^{s}	$	&	44	 & 	CC &[10,11,12]	\\
IKT~23	&	1E~0103.3-7240, DEM~S125, HFPK~217, SNR~0103--72.6, SNR~J0105.1--7223	&	$	01	^{h}	05	^{m}	04	^{s}	$	&	$	-72	^{d}	22	^{m}	56	^{s}	$	&	170	 & 	CC &[13,14]	\\
DEM~S128	&	SNR~0103--72.4, B0104--72.2, ITK~24, HFPK~145, SNR~J0105.4--7209	&	$	01	^{h}	05	^{m}	23	^{s}	$	&	$	-72	^{d}	09	^{m}	26	^{s}	$	&	124	 & 	Ia & [1,2]	\\
DEM~S130	&	SNR J0105.6-7204	&	$	01	^{h}	05	^{m}	39	^{s}	$	&	$	-72	^{d}	03	^{m}	41	^{s}	$	&	79	 & 	?& 	\\
IKT~25	&	B0104-72.3, HFPK125, SNR 0104-72.3, SNR J0106.2-7205	&	$	01	^{h}	06	^{m}	14	^{s}	$	&	$	-72	^{d}	05	^{m}	18	^{s}	$	&	110	 & 	Ia/CC &[1,2,15--19]	\\
N83C	&	NS83, SNR~J0114.0--7317,DEM~S147, SNR~J011333--731704, B0113=-729, SMC~B0112--7333	&	$	01	^{h}	14	^{m}	00	^{s}	$	&	$	-73	^{d}	17	^{m}	08	^{s}	$	&	17	 & 	?&	\\
            \bottomrule
\end{tabular}
}
\begin{flushleft}
[1] \citet{2004A&A...421.1031V}; [2] \citet{2015ApJ...803..106R}; [3] Auchettl et al. (2018), in prep. [4] \citet{2005ApJ...622L.117H}; [5]  \citet{2014ApJ...791...50S}; [6] \citet{haberl12}; [7] \citet{2011A&A...530A.132O}; [8] \citet{2015A&A...584A..41M}; [9] \citet{2002PASJ...54...53Y}; [10] \citet{1989ApJ...338..812B}; [11] \citet{2000ApJ...537..667B}; [12] \citep{2010ApJ...721..597V}; [13] \citet{2003ApJ...598L..95P}; [14] \citet{2002ApJ...564L..39P}; [15] \citet{2011ApJ...731L...8L}; [16] \citet{2014ApJ...788....5L}; [17] \citet{2007MNRAS.376.1793P}; [18] \citet{1994AJ....107.1363H}; [19] \cite{2016PASJ...68S...9T}
\end{flushleft}
\end{table*}

\section{Star formation history of the SMC}\label{sect2}

In this paper, we exploit the SFH map of the SMC from \citet{2004AJ....127.1531H}. Using UBVI stellar photometry of $\sim$6 million stars from the Magellanic Clouds Photometric Survey \citep{2002AJ....123..855Z},  \citet{2004AJ....127.1531H} derived color-magnitude diagrams in a rectilinear grid of 351 sub-cells that cover the central $4^{\circ}\times4.5^{\circ}$ of the SMC. Each subregion is $\sim12' \times 12'$ in size, which corresponds to approximately 200 pc $\times$200 pc in size at the distance of the SMC.  In each of these subregions, the local SFH  was derived using the SFH reconstruction program StarFISH \citep{2001ApJS..136...25H}. The SFH in each subregion was calculated for a lookback time from 4 Myr to 10 Gyr and was broken into three metallicity bins: $Z=$0.001, 0.004, and 0.008 (or [Fe/H]$=-1.3, -0.7$, and $-0.4$). 

\begin{figure*}[t!]
	\begin{center}
				\includegraphics[height=0.33\textwidth]{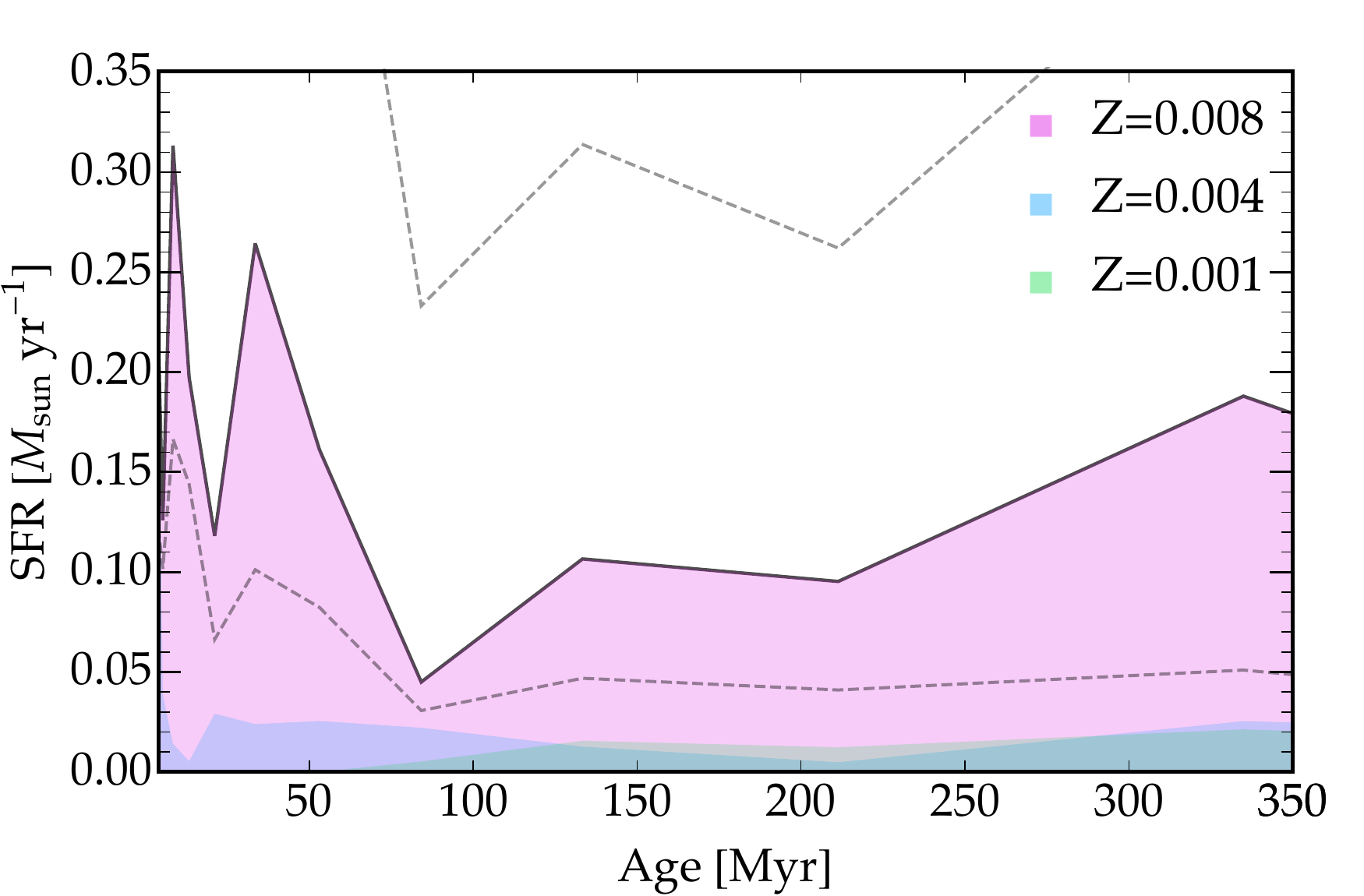}
		\includegraphics[height=0.33\textwidth]{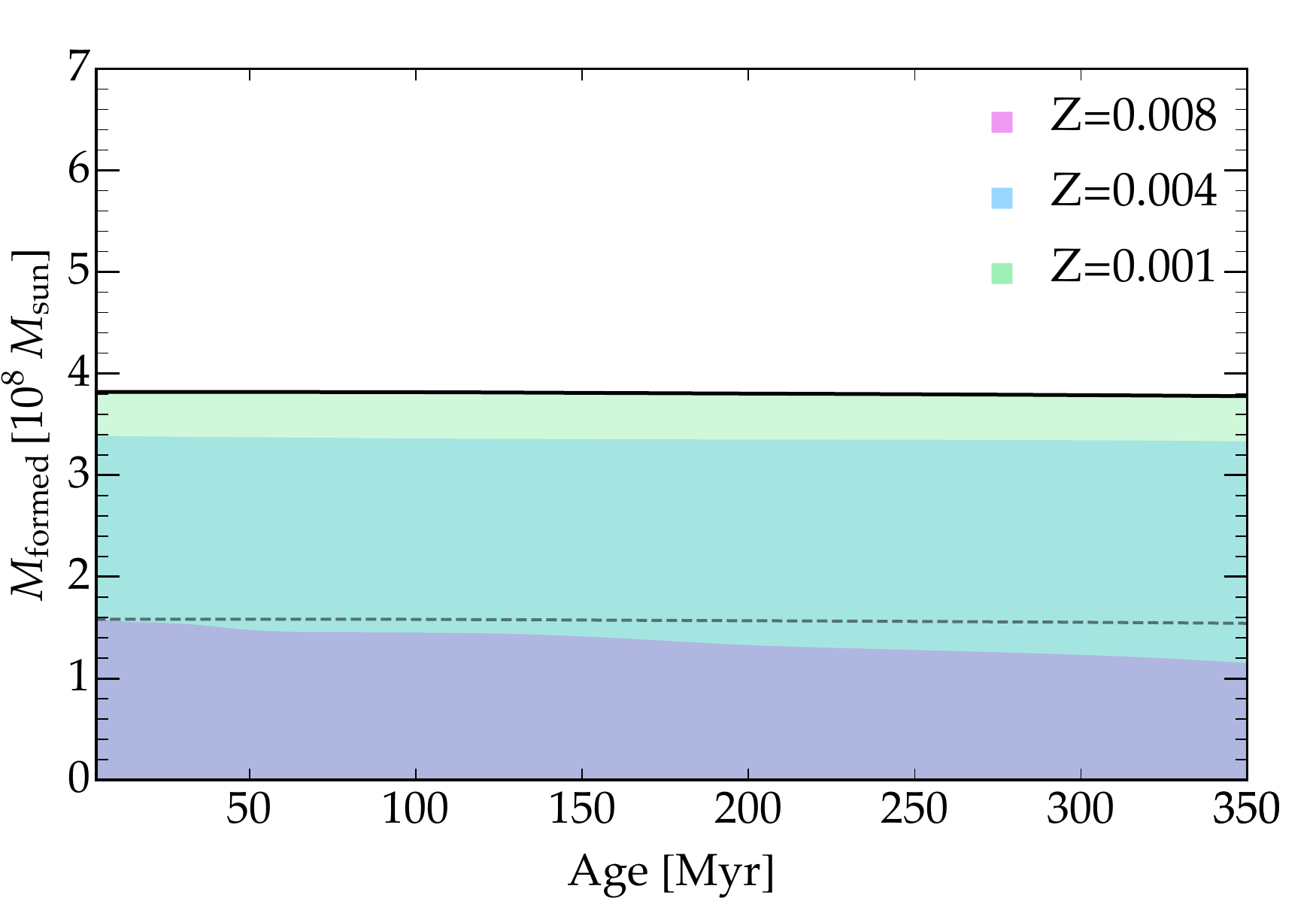}
		\includegraphics[height=0.33\textwidth]{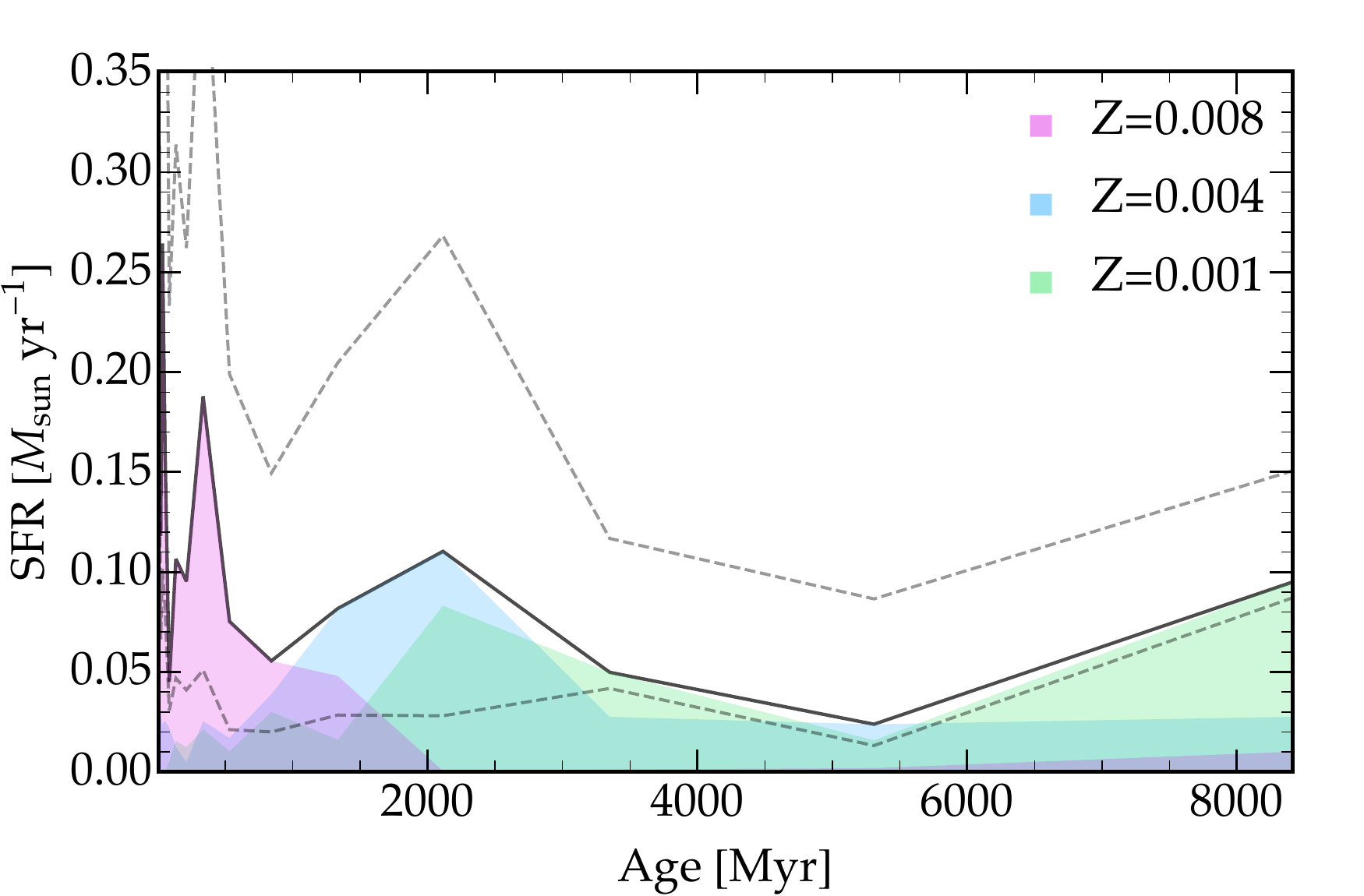}
		\includegraphics[height=0.33\textwidth]{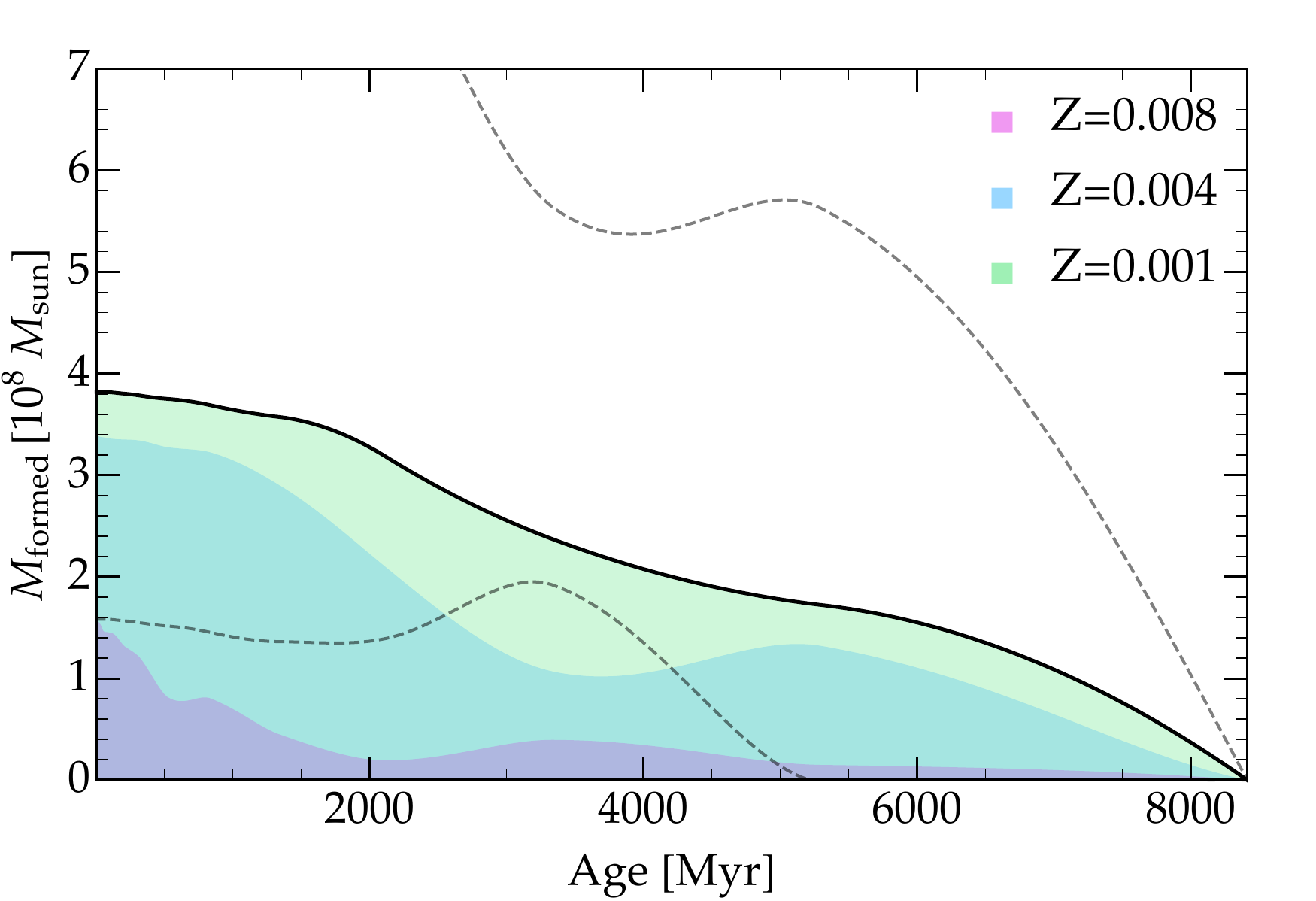}
		\caption{\textit{Left panels:} The total star formation rate (SFR) of the SMC as a function of look-back time as seen in the three metallicity bins defined in \citet{2004AJ....127.1531H}. The solid black line represents the total SFR, while the dashed grey lines represent the uncertainty on the total SFR. \textit{Right panels:} Cumulative stellar mass formed in the SMC as a function of look-back time as derived from integrating the SFR in each metallicity bin. We plot both the SFR and the cumulative stellar mass formed in the SMC over the timescales relevant for our analysis (top panels), and over the full look-back time of the SMC (bottom panels) for comparison.  \label{totalSFH}}
	\end{center}
\end{figure*}

In Figure~\ref{totalSFH} (right panels), we plot the total star formation rate (SFR) of the SMC as a function of lookback time over two different time ranges. The black solid line corresponds to the total SFH, and the dashed gray lines indicate the uncertainty in the total SFH. The pink, blue, and green curves indicate the SFR expected for the SMC assuming different metallicities ($Z=$0.008, 0.004, and 0.001, respectively). As discussed in detail in Section 2 and 3 of \citet{2004AJ....127.1531H}, the uncertainties in the SFR are most likely dominated by crowding effects. We note that studies of populations of B- and O-type stars found within the SMC \cite[e.g.,][]{2000A&A...353..655K, 2013A&A...555A...1B} have shown that $Z=0.004$ is representative of the stellar populations of the SMC. To be consistent with previous studies (such as \citealt{2009ApJ...700..727B}), we present all metallicities derived by \citet{2001ApJS..136...25H}. To avoid the effect of metallicity on our results, we use the total SFR to characterise the SFHs associated with our sources.

The SMC has a complex SFH, which most likely results from the fact that the SMC is gravitationally bound with the LMC and Milky Way \citep[e.g.,][]{2004ApJ...610L..93B, 2005MNRAS.356..680B,2005MNRAS.358.1215P,2016ARA&A..54..363D}. It is thought that the LMC and SMC underwent a recent and close encounter based on the detection of the Magellanic Bridge \citep{2012MNRAS.421.2109B}, proper motion measurements of the LMC and SMC \citep{2013ApJ...764..161K}, and the distribution of the OB stars in the MCs and the Magellanic Bridge \citep{2014ApJ...784L..37C}. As the SMC and LMC formed a binary pair approximately 2 Gyr ago \citep{2011MNRAS.413.2015D}, it is thought that this interaction triggered the enhanced star formation seen at lookback times $\gtrsim2$ Gyr as well as one more recently $\sim500$ Myr ago. However, further studies are needed to confirm that these close encounters are the origin of these burst epochs \citep{2016ARA&A..54..363D}.

Based on the SFH derived by \citet{2004AJ....127.1531H}, for lookback times $\gtrsim$8 Gyr ago, the SMC underwent significant SF in which approximately 50\% of all stars in the SMC were formed. Between 3 and 8.4 Gyr ago, the SMC underwent a period of quiescence, in which few stars were formed compared to other periods during its SFH. More recently ($<$3 Gyr), the SMC has been actively producing stars at a rate of $\sim0.1 M_{\odot}$ yr$^{-1}$, with bursts of SF around 2.5 Gyr, 400 Myr and 60 Myr ago. 

Figure \ref{totalSFH} (left panels) also highlights the evolution of the SMC's chemical enrichment. For a lookback time of $\gtrsim$3--5 Gyr ago, the SMC exhibited little SF and thus very little variation in its metallicity during this early period \citep[e.g.,][]{1991IAUS..148..393D, 1998AJ....115.1934D, 1999Ap&SS.265..461P, 2001MNRAS.325..792P}. However, when the SMC experienced two periods of enhanced SFH at $\sim$1--3 Gyr and $\sim$5--8 Gyr ago, the mean metallicity of the SMC increased from [Fe/H]$\sim-1.3$ to a metallicity close to that of the present day [Fe/H]$\sim-0.6$ \citep{1999Ap&SS.265..461P, 2001MNRAS.325..792P, 2001ApJ...562..303D}. 

This relatively uniform increase in metallicity is likely responsible for the fact that the SMC does not exhibit the "abundance or age" gap observed for its star clusters, as seen for those in the LMC \citep[e.g.,][]{1991AJ....101..515O}. As a consequence, the more metal-rich stars detected in the SMC are also the youngest, whereas the metal-poor stars are older and were formed $\gtrsim$ few Gyr ago. This behaviour is also seen in Figure \ref{totalSFH} (right panels), where we have plotted the cumulative stellar mass formed in the SMC as a function of lookback time. Here, most of the metal-poor stars have an age $\gtrsim$0.5--1 Gyr, and the more metal-rich stars tend to be $\lesssim$10-100 Myr.

\section{Source selection and the connection between Supernova Remnants and their supernova explosions. }\label{sect3}
 
\subsection{Source selection}

We characterise the SFHs at the sites of all SNRs in the SMC, regardless of age, because a large number of these sources have not been typed before. Given that SNRs arise from explosions that occurred $\sim1000$s of years ago, it can be difficult to discriminate the class of SNe that led to each SNR. Typically, SNRs are typed based on their metal abundances measured using X-ray spectra \citep[e.g.,][]{2012A&ARv..20...49V} or by detecting a compact object associated with the source, or through their X-ray morphologies \citep{2009ApJ...706L.106L, 2009ApJ...691..875L, 2011ApJ...732..114L,2014ApJ...781L..26G}. \citet{2009ApJ...700..727B} aimed to minimise the chances of mis-identifying the LMC SNR progenitors by limiting their sample to only the youngest (and smallest) SNRs and those with well-determined SN types in the literature. Thus, they considered 8 out of the 59 known SNRs in the LMC. By contrast, we examine the full population of SNRs in the SMC. In doing so, we aim to confirm the nature of the well-studied or controversial SNRs in the SMC as well as to set initial constraints on those without any SN classification in the literature.

\subsection{Progenitors of CC and Type Ia SNe} \label{progenitors}

To characterise the progenitor masses of CC SNe, we adopted the single-star models of \citet{2017arXiv170107032Z}. While these models are based on rapid stellar evolution algorithms, they provide similar results to detailed stellar models from \citet{2004MNRAS.353...87E} and \citet{2008MNRAS.384.1109E} that were used in \citet{2009ApJ...700..727B}. Theoretically, these models predict that Type IIP SNe arise from 8--30$M_{\odot}$ red supergiants with their full H envelopes intact, Type IIb SNe come from 30--40$M_{\odot}$ red supergiants with a partial H envelopes, and Type Ib/Ic result from stars $>$40$M_{\odot}$ that lose all of their H envelopes\footnote{We note that Type Ib/Ic SNe can be bipolar/jet-driven. In fact, as rapid rotation without extensive mass loss is required to produce bipolar SNe, it expected that these explosions will be found predominantly in low-metallicity environments like the SMC \citep[e.g.,][]{2002ApJ...565L...9R, 2004MNRAS.348.1215I}.},\footnote{The discovery of Type Ib supernova iPTF13bvn by \citet{2013ApJ...775L...7C} has brought into question whether Type Ib SNe result predominantly from progenitors $>40M_{\odot}$. Work by e.g., \citet{2014AJ....148...68B} and \citet{2015MNRAS.446.2689E} have shown that this explosion likely arose from a low-mass progenitor  ($\sim$11~$M_{\sun}$)  found in a binary system.}.  In these models, the lifetime of a 8$M_{\odot}$ star is $\sim$40 Myr, whereas the lifetime of a $>$40$M_{\odot}$ star is $\lesssim$ 5 Myr.

Observationally, the sub-type of a SN is determined by early time spectra of these events. However, characterising SN progenitor masses is difficult as it relies on the existence of pre-explosion images, and only a handful of sources have been observed prior to the SNe. \citet{2009MNRAS.395.1409S} analyzed pre-SN explosion images and showed that known Type IIP/IIb SNe have progenitor masses of 8.5--16.5$M_{\odot}$, suggesting a lack of high-mass red supergiant progenitors above 17$M_{\odot}$. It is possible that this result arises from systematics related to under-estimating progenitor masses due to extinction, or that stars $>$17$M_{\odot}$ are produced by other SN types, such as Type II-L/II-n/II-b. However,  \citet{2009MNRAS.395.1409S} suggest that neither of these explanations is ideal, and thus it is likely that more massive progenitors possibly formed black holes with faint or non-existent SNe. This conclusion was also reached by \citet{2013MNRAS.436..774E}, who attempted to place constraints on the progenitor masses of Type Ib/Ic SNe using the low ejecta masses derived from the literature \citep[see review by][]{2015PASA...32...16S} and upper limits from pre-explosion images. These authors noted that the rates of Type Ib/Ic SNe are quite high and can only be achieved if Type Ib/Ic SNe result from a mixed population of single-stars and binary systems with stars of initial mass $<$20$M_{\odot}$.

Generally, it is difficult to distinguish the CC SN subtype of SNRs. Only a handful of SNRs have well-constrained CC SN subtypes derived using X-ray, infrared, or optical  observations\footnote{For example, Cassiopeia~A is thought to have been from a Type IIb explosion based on the expansion properties of its shock-front \citep{2003ApJ...593L..23C, 2005ApJ...619..839C} and light echo spectra \citep{2008Sci...320.1195K}. See review by \citet{2017arXiv170100891M} for a more detailed overview.}. Due to the challenges of subtyping SNRs, our study aims to ascertain whether individual SNRs likely arose from a CC explosion, based on a recent burst of star formation, and the probable progenitor mass using the massive star lifetimes of \cite{2017arXiv170107032Z}, rather than focusing on what subtype of explosion it underwent.

It has been shown that massive O stars are found in binary systems. Of these, more than 70\% will have their evolution affected by binary interaction, with approximately one third of these massive stars the result of a binary merger \citep{2012Sci...337..444S}. Recently, \citet{2017arXiv170107032Z} demonstrated that $\sim$15\% of massive stars found in these binary systems will undergo a delayed CC SN, exploding after $\sim$50--200 Myr, in contrast to the  $\sim$3--50 Myr expected for a single massive star. Consequently, star formation at these lookback times may suggest a CC SN of a close binary system or a ``prompt'' Type  Ia SN origin, which is discussed below.

A natural consequence of binary interaction is that at least $\sim20$\% of massive main-sequence stars will be rapidly rotating \citep[e.g.,][]{2013ApJ...764..166D}. Rapid rotation increases mixing within stars, extending the lifetimes and evolutions of those stars \citep[][]{2000ARA&A..38..143M}. In low-metallicity environments, the winds of massive main-sequence stars are weak \citep[e.g.,][]{2001A&A...369..574V, 2007A&A...473..603M}. As stars lose their angular momentum via stellar winds \citep{1998A&A...329..551L}, stars in these environments will lose less angular momentum and thus spin much faster than those found in solar-metallicity environments \citep[e.g.,][]{2007A&A...462..683M, 2015MNRAS.453.2637D, 2016MNRAS.458.4368M, 2017MNRAS.465.4795B, 2017ApJ...846L...1D}. In fact, \citet{2008A&A...479..541H, 2009A&A...504..211H} measured the rotational velocities of a large population of massive stars in both the Milky Way and Magenallic Clouds and found that massive stars in the SMC rotate faster than those in the Milky Way with a 3$\sigma$ confidence. 

\begin{figure}[t!]
	\begin{center}
		\includegraphics[width=\columnwidth]{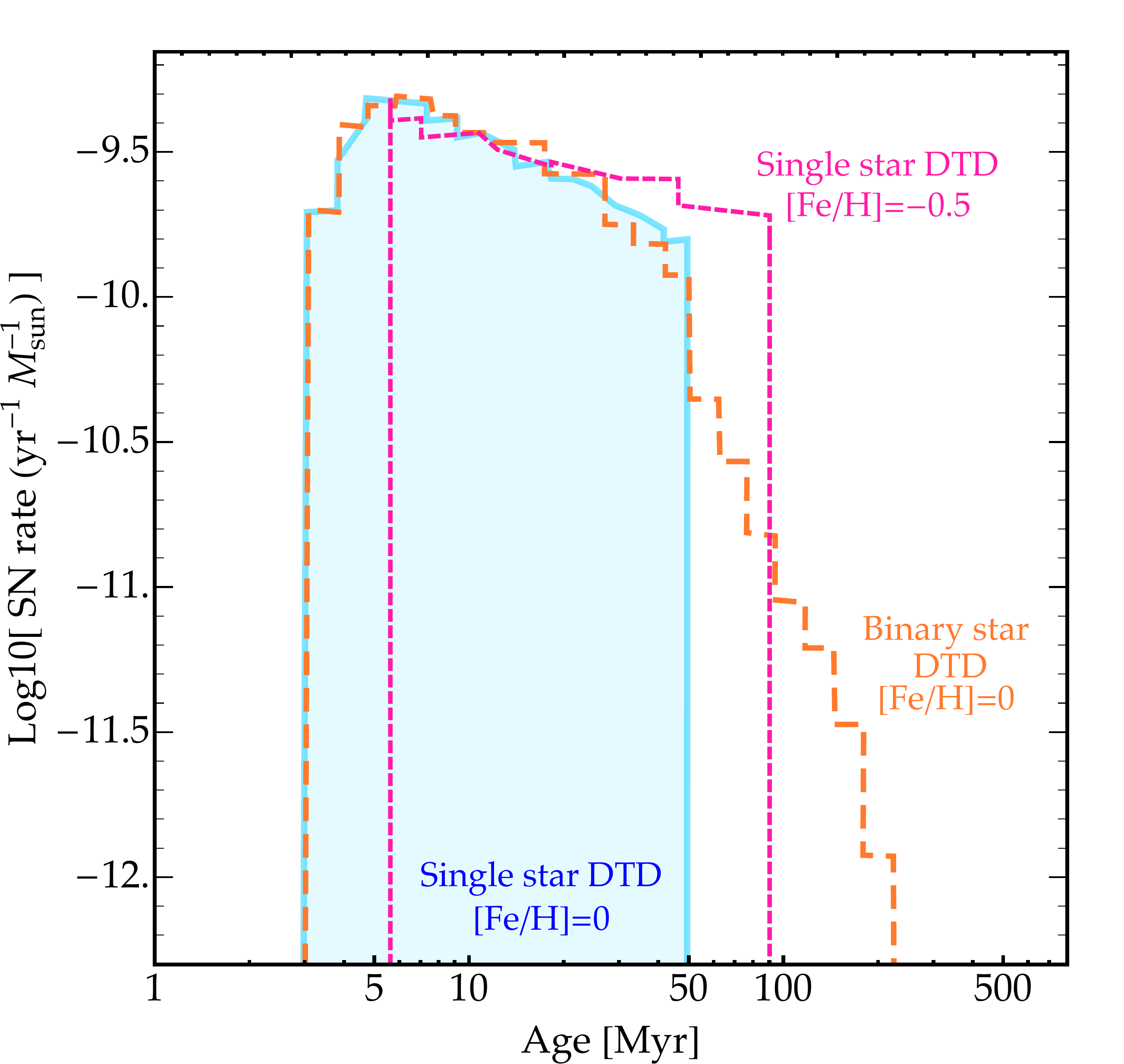}
		\caption{The effect of binarity (orange dashed curve) and rapid rotation (magenta dotted curve) on the delay-time distribution (DTD) of CC SNe arising from a single-star population (blue shaded). Here the solar metallicity ([Fe/H]=0) single-star and binary star DTD are adapted from \citet{2017arXiv170107032Z}, while we use the MESA isochrone and stellar track of \citet{2016ApJ...823..102C} which assumes that all stars rapidly rotate to quantify the effect that rapid rotation in a low metallicity environment ([Fe/H]=$-$0.5) has on the lifetimes, and thus the DTD, assuming a single-star population. \label{compare}}
	\end{center}
\end{figure}

In Figure~\ref{compare}, we plot how binarity\footnote{Here and throughout the text we define binarity as the effect of binary interactions on a stellar populations.} and rapid rotation alters the delay-time distribution (DTD) of CC SNe arising from a single-star population.	 Plotted as the blue solid and orange dashed curve are the solar metallicity ([Fe/H]=0) single-star and binary-star DTD, respectively, from \citet{2017arXiv170107032Z}. One can see that binarity leads to a fraction of progenitors that will undergo CC well after all massive single-stars have already exploded.

To illustrate the effect that rapid rotation can have on the lifetimes of massive single-star populations, we use the MESA isochrone and stellar tracks of \citet{2016ApJ...823..102C}, which assumes that all stars are rapidly rotating. Using the same form of the DTD as for a single-star population at solar metallicity, we have plotted the corresponding DTD (the magenta dotted curve in Figure~\ref{compare}), assuming a metallicity similar to that of the SMC ([Fe/H]=$-$0.5). One can see that the lifetime for the last SN assuming a single-star population can be significantly delayed in the rapidly rotating, low-metallicity case. However, we note that not all massive stars in the SMC are rapidly rotating \citep{2008A&A...479..541H, 2009A&A...504..211H}, and thus the extent of this delay is likely an upper-limit for when the last CC SN would explode.

As the SMC is a low-metallicity galaxy, we also note the effect that metallicity, independent of rotation, has on the lifetimes of SNe. Recently, using their suite of binary stellar evolution models which span a wide range of masses and metallicities, \citet{2017PASA...34...58E} investigated the effect that metallicity alone has on the DTDs of single-star and binary star populations. They found that the effect of metallicity on the age distribution of SNe arising from a single-star population is relatively minor, while for binary populations, there is a small increase in the lifetimes \citep[see Figure 19 of ][]{2017PASA...34...58E}.

For Type Ia SNe, the identity of their progenitors and the mechanism responsible for their explosion remains uncertain both theoretically and observationally \citep[see e.g.,][]{2014ARA&A..52..107M, 2016IJMPD..2530024M}. These events are thought to occur when either the mass accreted onto the WD comes close to the Chandrasekhar limit \citep[e.g.,][]{1996ApJ...470L..97H, 2004MNRAS.350.1301H} or the objects merge or collide on the dynamical timescales of these systems \citep[e.g.,][]{1984ApJS...54..335I, 2010ApJ...709L..64G, 2010ApJ...714L..52S, 2010ApJ...722L.157V,2011ApJ...737...89D}.

Significant work, both observationally and theoretically, has been done to explore the single degenerate and double degenerate scenarios of Type Ia explosions (see the review by \citealt[][]{2014ARA&A..52..107M}, and more recent work by e.g., \citealt{2015ApJ...801L..31Y} and \citealt{2016ApJ...825...57M}). In terms of SFHs, Type Ia SNe have been shown to be associated with galaxies that exhibit a variety of SFHs and galaxy properties \citep[e.g.,][]{2008A&A...492..631A, 2010MNRAS.407.1314M, 2011MNRAS.412.1441L, 2017ApJ...848...25M}. 

As a consequence, Type Ia SNe arise from a wide range of progenitors of various ages, with the rate of Type Ia SN detected well described using a simple powerlaw (also called the delay time distribution, see Section \ref{sect4} for more details). As such, the young (``prompt'') progenitor population that produces Type Ia SNe on timescales of  $\lesssim100-330$ Myr \citep[e.g.,][and references therein]{2008A&A...492..631A, 2010MNRAS.407.1314M} have a much greater representation than the older (``delayed") progenitor population which explode on timescales of $\sim$Gyr and are not associated with star formation.

 \section{Local star formation history around individual Supernova remnants}\label{sect4}

In disk-dominated galaxies, the mixing of stellar populations due radial migration from the gravitational interaction of stellar populations results in the SFHs directly surrounding a SNR to be representative of the SN progenitor up to a specific lookback time \citep[e.g.,][]{2002MNRAS.336..785S, 2012MNRAS.426.2089R, 2013A&A...553A.102D}. However, for dwarf galaxies which lack a well defined disk, bar or spiral arm, simulations have shown that stellar mixing and migration only becomes important (if at all) on cosmological timescales \citep[e.g.,][and references therein]{2016ApJ...820..131E}. As a consequence, the stellar populations associated with these galaxies change very little over the lifetime of the host galaxy \citep[e.g.,][]{2009MNRAS.395.1455S, 2013MNRAS.434..888S}. The SMC, which is an irregular dwarf galaxy, does not exhibit a well defined disk or spiral structure \citep[e.g.,][]{2004ApJ...604..176S}. As a result, stellar mixing and migration is likely not an important effect in the SMC, and thus the stellar populations surrounding each of the SNRs should be representative of the SN progenitors. For CC SNe, this conclusion is also supported by work done by \citet{2008MNRAS.390.1527A}, who showed that the overall population of CC SNe follows closely H$\alpha$ emission, an excellent tracer of recent star formation, with SNe that result from high-mass progenitors (such as SNe Ib/c) in regions of strong H$\alpha$ emission (see Figure 1). 

Additionally, even though most massive stars are born in a binary system, a majority of these systems are disrupted after one of the companions undergoes CC. Then these unbounded, young massive stars can gain kick velocities of a few km~s$^{-1}$, becoming what is known as walkway stars \citep{2012ASPC..465...65D}, while a small fraction of these companions become runaway stars with velocities $>$30 km~s$^{-1}$ \citep{2011MNRAS.414.3501E, 2018arXiv180409164R}. For stars $>15M_{\odot}$, $\sim$10\% will have velocities consistent with a walkaway star, while $\sim$0.5\% will have velocities $>$30 km~s$^{-1}$ \citep{2018arXiv180409164R}.

In low-metallicity environments like the SMC, runaway stars tend to be more common as they experience less mass loss, and thus they are found in systems that are more difficult to disrupt without strong kicks.  \citet{2018arXiv180409164R} showed that both walkaway and runaway stars with masses $>7.5M_{\odot}$ travel a mean distance of $\sim$163 pc in a low-metallicity environment before exploding as supernovae, and more massive progenitors travel substantially less  \citep{2011MNRAS.414.3501E, 2018arXiv180409164R}. Since the size of the subregions used in our analysis is $\sim$200 pc, only the fastest runaway stars would drift significantly from their birth place. However, as these very fast objects are rare, it is unlikely that they contribute significantly to the SFH measured in the proximity of each SNR. As such, the stellar populations surrounding each remnant should be representative of the SN progenitors.

\begin{figure*}[ht]
	\begin{adjustbox}{addcode={\begin{minipage}{\width}}{\caption{%
						The local star formation histories of the SMC SNRs as a function of lookback time up to 150 Myrs in three metallicity bins. The coloured, shaded bands correspond to the one-$\sigma$ uncertainty in the SFR as determined by \citet{2004AJ....127.1531H}. The dashed vertical line is the lifetime of an isolated 8$M_{\odot}$ star from the models of  \cite{2017arXiv170107032Z} . \label{SFHfull1}
			}\end{minipage}},rotate=90,center}
		\includegraphics[width=\textheight]{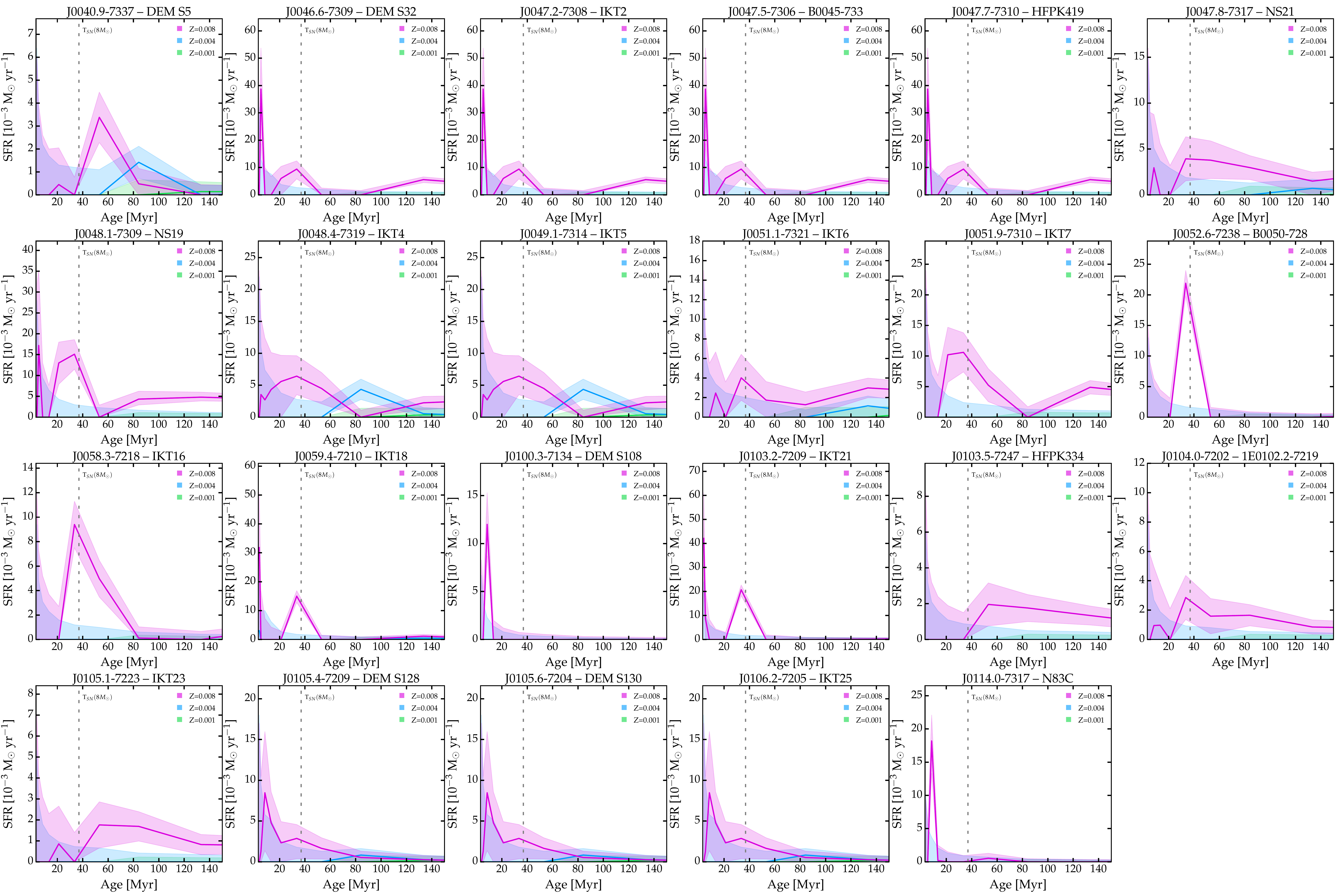}%
	\end{adjustbox}
\end{figure*}

\begin{figure*}[ht]
	\begin{adjustbox}{addcode={\begin{minipage}{\width}}{\caption{%
						The recent SFHs in the vicinities of the SMC SNRs. Here we plot only the SFH associated with a metallicity of $Z=0.008$, comparable to the measured metallicity of the SMC now. We have binned the SFHs into three-mass intervals: 8--12.5\,$M_{\odot}$, 12.5--21.5\,$M_{\odot}$ and $>21.5M_{\odot}$. We have included the 50--200 Myr time range as well to identify possible delayed CC SNe due to either the low-metallicity environment or to binary interaction. The range at the top of the figures correspond to the zero-age main-sequence progenitor masses for different ages as derived from the single-star models of \cite{2017arXiv170107032Z}. \label{SFH1}
			}\end{minipage}},rotate=90,center}
		\includegraphics[width=\textheight]{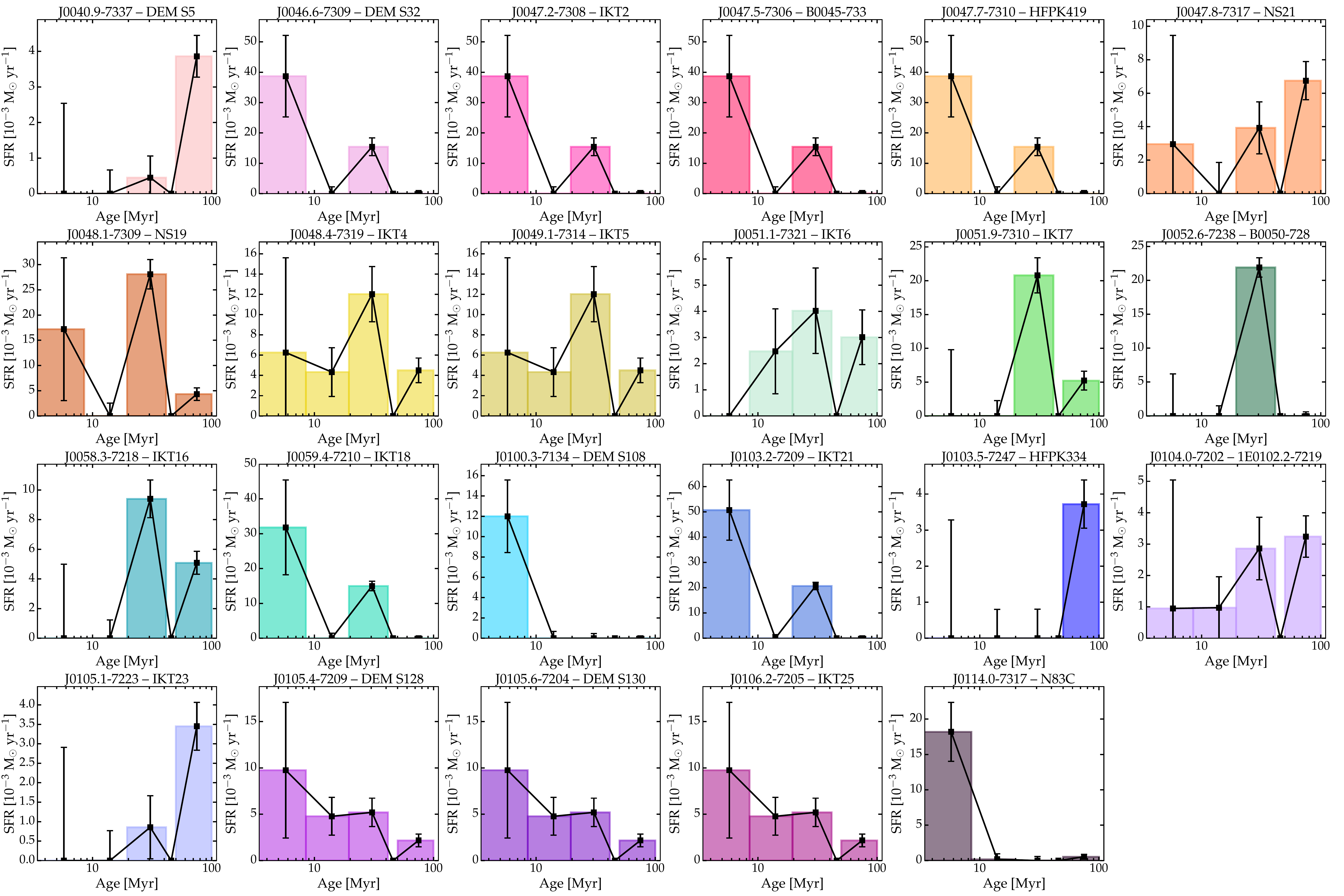}%
	\end{adjustbox}
\end{figure*}

In Figure \ref{SFHfull1}, we plot the local SFHs as a function of lookback time for each SNR in the SMC, using the subcell that is coincident with each SNR. We have overlaid a dashed vertical line to denote the lifetime of an isolated 8$M_{\odot}$ star assuming solar metallicity \citep{2017arXiv170107032Z}. Here we focus on the most recent time bins that are relevant for progenitor identification. Figure \ref{SFH1} shows the same SFHs for each SNR as in Figure \ref{SFHfull1}, but we have binned the SFH into three time intervals corresponding to the progenitor mass ranges of  8--12.5\,$M_{\odot}$, 12.5--21.5\,$M_{\odot}$, and $>21.5M_{\odot}$. Additionally, we have included the 50--200 Myr time bin to aid in identification of massive progenitors that may have undergone delayed CC due to binary interaction and rapid rotation \citep[e.g.,][]{2013ApJ...764..166D, 2015ApJ...805...20S, 2017arXiv170107032Z} or the low metallicity \citep[e.g.,][]{1998MNRAS.298..525P} of the SMC. Overlaid at the top of each plot in Figure \ref{SFH1} is the time when stars of a given zero-age main-sequence (ZAMS) mass M$_{\rm ZAMS}$ would have formed, assuming the single-star models of \cite{2017arXiv170107032Z}. 

To make statistical statements in regards to the progenitor distributions implied by the local SFHs near SNRs, we take advantage of current delay-time distributions (DTDs) associated with both CC and Type Ia SNe.  The DTD is an important tool used in stellar population studies as it encapsulates the timescales in which members of a particular stellar population born in a burst of star formation will go undergo a SN \citep[e.g.,][]{2012PASA...29..447M,2014ARA&A..52..107M}. Combined with the local SFH, the DTD allows us to constrain the progenitor origin of each SNR, since the form of the DTD is set by the progenitor population and by the stellar and binary evolution of that population \citep[e.g.,][]{2012PASA...29..447M, 2014ARA&A..52..107M, 2017ApJ...848...25M,2017arXiv170107032Z}.

As Type Ia SNe originate from lower-mass systems than CC SNe, the former tend to have delay times on the order of Myr to Gyr. As such, most studies related to observationally constraining the DTDs of SNe have focused on quantifying the Type Ia DTD using extragalactic SN surveys \citep[see review by][]{2014ARA&A..52..107M}. However, CC SNe (which arise from more massive progenitor systems) have delay times that are significantly shorter, on timescales of Myr. Consequently, it is much more difficult to observationally constrain these timescales following the same techniques used for Type Ia SN \citep[e.g.,][]{2010MNRAS.407.1314M, 2011MNRAS.412.1508M}, since the DTD of CC SNe requires a much more accurate determination of the ages of the underlying stellar population to derive this distribution.

Significant theoretical effort has gone into determining the DTD of CC SNe. \citet{2017arXiv170107032Z} performed a detailed population synthesis study to derive the DTD of CC SNe, taking into account that 70\% of massive stars evolve in binary systems \citep[e.g.,][and references therein]{2012Sci...337..444S}. Compared to the DTD of CC SNe assuming a single-star population which cuts off after the last single-star explodes ($\sim$50~Myr), \citet{2017arXiv170107032Z} find that binarity tends to extend the DTD to longer delay times (50--200 Myr). These delayed events represent a non-negligible contribution ($\sim15.5\%$) to the total number of CC SNe assuming a Kroupa initial mass function (IMF). In fact, the enhanced SN rate between 35--330 Myr seen in the LMC and SMC by \citet{2010MNRAS.407.1314M} could suggest a substantial contribution of delayed CC SNe.

Due to the importance of using DTDs to guide our conclusions about the nature and progenitor properties of each SNR, we convolve our local SFHs (Figure \ref{SFH1}) with recent DTDs for both Type Ia and CC SNe. For our study, we use the Type Ia DTD derived by \citet{2017ApJ...848...25M}, which has the form of $t^{-(1.1\pm0.1)}$ and a total integrated rate of  $N/M_{\star} = (1.3\pm0.1)\times10^{-3}~M_{\odot}^{-1}$.  For CC SNe, we use both the single-star and binary star population DTDs derived by \citet{2017arXiv170107032Z} to quantify the masses of the progenitors and to test whether any of the SMC SNRs could have resulted from a delayed CC explosion.

By correlating the timing of the various star formation bursts, we can estimate the mass of the progenitor and associate a likelihood that the SNR results from a particular SN subtype. To estimate the likelihood, we convolve the three most recent SFH bins that are less than 50 Myr with a standard Salpeter initial mass function \citep{1955ApJ...121..161S} to determine the likelihood that the progenitors in one of the three mass intervals results in a CC SNe.

For sources thought to arise from Type Ia SNe, we quantify the fraction of the stellar population that would result from the prompt and delayed channels by assuming that the rate of Type Ia SNe ($R_{\rm Ia}$) in a galaxy is given by $R_{\rm Ia}=R_{\rm delayed} + R_{\rm prompt}$. Here we assume that prompt and delayed Type Ia SN arise on timescales of $\sim100$ Myr and $>2.4$ Gyr, respectively \citep{2012MNRAS.426.3282M}, and convolve the local SFR within these time frames with the Type Ia DTD of \citet{2017ApJ...848...25M}.

Using the total SFR and the cumulative stellar mass plots presented in Figure~\ref{totalSFH} (bottom panels), we can calculate $R_{\rm delayed}$ and $R_{\rm prompt}$ for the whole SMC: $R_{\rm delayed}=10\%$ and $R_{\rm prompt}=90\%$. This is quite different to that derived for the LMC: $R_{\rm delayed}=41\%$ and $R_{\rm prompt}=59\%$ \citep{2009ApJ...700..727B}.  However this is not so surprising since the rate of SNe per year used by \citet{2009ApJ...700..727B} is an order of magnitude lower than the SN-Ia rate derived from recent observations \citep[e.g.,][]{2017ApJ...848...25M}. Using the Type Ia SN rate of \citet{2006ApJ...648..868S} (which was adopted by \citet{2009ApJ...700..727B} in their calculation), we find that $R_{\rm delayed}=33\%$ and $R_{\rm prompt}=67\%$, which is still lower than that derived for the SMC. One possible reason for this disparity between the SMC and LMC is their different metallicities. In particular, it has been shown by \citet{2013ApJ...770...88K} that the mass of a carbon-oxygen white dwarf increases for lower metallicities. The stars that produce these more massive white dwarfs would evolve much more rapidly compared to those found in high metallicity environments, which could lead to the prompt channel dominating the Type Ia SNe in low-metallicity galaxies.

In the following subsections, we provide a general overview of the results we obtained by considering both the SFHs around each SNRs with the properties of the SNR and its environment derived from X-ray/optical observations. A more detailed description of these properties and SFHs of each of the SMC remnant can be found in the Appendix. Table~\ref{CCSNRs} summarizes the possible progenitor type of each SNR based on the SFHs and the likelihood of a particular CC progenitor or prompt/delayed Type Ia progenitor.

\subsection{Core-Collapse SNRs} \label{cc}

Of the 23 SNRs in the SMC, only five of these sources (HFPK~419 [SNR J0047.7$-$7310], IKT~6 [B0049$-$73.6], IKT~16 [SNR J0058.3$-$7218], 1E~0102.2--7219 [SNR B0102$-$7219] and IKT~23 [SNR 0103$-$72.6]) have been classified robustly as arising from a CC explosion. Five other remnants, DEM~S32 [SNR J0046.6$-$7309], IKT~2 [B0045$-$73.4], IKT~7 [SNR J0051.9$-$7310], IKT~18 [SNR J0059.4$-$7210] and IKT~21 [B0101$-$72.4], have also been suggested to be CC SNRs in the literature, while IKT~5 [SNR 0047-73.5] and IKT~25 [SNR 0104-72.3] have been claimed as Type Ia or as CC SNRs by different authors. Among these 12 SNRs, five have progenitor mass estimates in the literature, and we compare our results below to test the robustness of the SFHs as tools to estimate progenitor masses.

The most striking feature of the SFHs of the stellar populations associated with a majority of these SNRs is the strong burst of SF seen at times $<50$ Myr, similar to the CC LMC SNR sample presented in \citet{2009ApJ...700..727B}. However, some variation in the SFHs between sources is evident. For example, we find that for IKT~25, the SFR burst is less than that seen around e.g., HFPK~419 and IKT~16. For IKT~6 and 1E~0102.2--7219, their associated stellar populations do not exhibit any bursts in SFR, instead showing persistent SF that increased 10--100 Myr ago. 

In addition to their burst of SF $<50$ Myr ago, we also note that the stellar populations associated with IKT~6, IKT~16, 1E~0102.2--7219, IKT~23, IKT~7, IKT~5, and IKT~25 exhibit a strong peak of SF in the 50--200 Myr bin. This burst of SF suggests that some of the stellar population associated with these remnants may have undergone a delayed CC either due to binary interaction, rapid rotation or low metallicity. Assuming a binary star population DTD, it seems that for a majority of them, no more than $\sim$10\% of the stellar population underwent delayed CC. However, for IKT~23 and IKT~25, our analysis suggests that a significant fraction ($\sim42$\% and $\sim22$\%, respectively) of their stellar population underwent delayed CC.

For IKT~6, 1E~0102.2--7219, IKT~2 and IKT~25 which have independent progenitor mass estimates from detailed X-ray or optical studies, the SFRs of their stellar populations $<50$ Myrs ago are consistent with the lifetimes of the progenitor masses implied from these studies. This suggests that for a majority of known CC remnants the SFH can provide a relatively robust and independent confirmation of progenitor masses for these systems. The exception to this is IKT~23 whose SFR $<$50 Myr ago implies that 100\% of the CC SN progenitors associated with this remnant have masses of 8--12.5$M_{\odot}$, which is slightly below the $\sim18M_{\odot}$ suggested by \citet{2003ApJ...598L..95P}. 

The stellar populations associated with IKT~16,  IKT~2, 1E~0102.2--7219 and IKT~7 exhibit a strong burst of SF consistent with either a $8-12.5M_{\odot}$, and a $>21.5M_{\odot}$ progenitor. The progenitor masses implied by the SFHs associated with these remnants is consistent with theoretical work by e.g., \citet{2016ApJ...821...38S} who suggested that neutron stars are produced in SNe of stars $<30M_{\odot}$. These four remnants have been suggested to harbour either a PWN/CCO or is associated with a Be/X-ray pulsar system \citep{2000PASJ...52L..37Y, 2006AAS...20915614W, 2011A&A...530A.132O, haberl12, 2015A&A...584A..41M, 2018arXiv180301006V}. 

The stellar populations associated with the three oxygen rich remnants, IKT~6, 1E~0102.2--7219, and IKT~23, have some similarities in their SFH $<50$ Myrs ago. Our method is unable to constrain whether the oxygen-rich SNRs are associated with a particular progenitor mass. We should note that all three show significant SF between 50--200 Myr ago, however there is also a number of other remnants which are not oxygen rich and do show this burst at much later times.

Out of the 12 CC SNRs discussed above, eight of them (HFP~419, 1E~0102.2--7219, DEMS~32, IKT~2, IKT~18, IKT~21, IKT~5 and IKT~25) have SFHs that suggest a large fraction of CC SN progenitors associated with the stellar population of each remnant have a mass $>21.5M_{\odot}$. The other four remnants (IKT~6, IKT~16, IKT~23, and IKT~7), have SFHs consistent with a progenitor $<21.5M_{\odot}$. This result will be discussed further in Section \ref{sect5}.

\begin{table*}[t!]
	\centering
	\caption{SMC SNR classification based on SFH and the likelihood of their progenitor properties.}
	\label{CCSNRs}
\resizebox{\textwidth}{!}{
	\begin{tabular}{p{1.85cm}ccccccccccccccccccccc}
		\hline
				   & &  &  \multicolumn{4}{c}{Assuming single star population} &&  \multicolumn{5}{c}{Assuming binary star population} & \multicolumn{2}{c}{  }\\ 
				   \cline{4-7} \cline{9-13}\\
		Name   &SNe origin& Evidence of&  $N_{\rm CC}/N_{\rm Ia}$&\multicolumn{3}{c}{\% of CC SNe progenitors from} & &  $N_{\rm CC}/N_{\rm Ia}$&\multicolumn{4}{c}{\% of CC SNe progenitors from} & \multicolumn{2}{c}{\% of Type Ia progenitors from }\\ 
		
		&  based on SFHs &delayed CC?&&$>21.5  M_{\odot}$ & $12.5-21.5  M_{\odot}$&$8-12.5  M_{\odot}$ &&&$>21.5  M_{\odot}$ & $12.5-21.5  M_{\odot}$&$8-12.5  M_{\odot}$ & delayed CC& prompt & delayed \\ 
		\hline
		\hline
DEM~S5 	&	CC$^{\dagger}$	&	d	&	3	&	0	&	0	&	100	&	&	3	&	0	&	0	&	44	&	56	&	95	&	5	\\
DEM~S32 	&	CC$^{\dagger}$ 	&	$\dotsm$	&	14	&	92	&	0	&	8	&	&	15	&	93	&	0	&	7	&	0	&	$\dotsm$	&	$\dotsm$	\\
IKT~2 	&	CC$^{\dagger}$	&	$\dotsm$	&	14	&	92	&	0	&	8	&	&	15	&	93	&	0	&	7	&	0	&	$\dotsm$	&	$\dotsm$	\\
B0045--733 	&	CC$^{\dagger}$ 	&	$\dotsm$	&	14	&	92	&	0	&	8	&	&	15	&	93	&	0	&	7	&	0	&	$\dotsm$	&	$\dotsm$	\\
HFPK~419 	&	 CC*$^{\dagger}$	&	$\dotsm$	&	14	&	92	&	0	&	8	&	&	15	&	93	&	0	&	7	&	0	&	$\dotsm$	&	$\dotsm$	\\
NS~21 	&	CC$^{\dagger}$	&	(d?) 	&	3	&	78	&	0	&	22	&	&	3	&	75	&	0	&	19	&	6	&	99	&	1	\\
NS~19 	&	CC$^{\dagger}$	&	(d?)	&	11	&	74	&	0	&	26	&	&	12	&	75	&	0	&	24	&	1	&	$\dotsm$	&	$\dotsm$	\\
IKT~4 	&	CC	&	(d?) 	&	7	&	59	&	17	&	24	&	&	7	&	58	&	18	&	22	&	2	&	98	&	2	\\
IKT~5 	&	CC$^{\dagger}$	&	(d?)	&	7	&	59	&	17	&	24	&	&	7	&	58	&	18	&	22	&	2	&	$\dotsm$	&	$\dotsm$	\\
IKT~6 	&	CC*	&	(d?) 	&	4	&	0	&	54	&	46	&	&	4	&	0	&	53	&	39	&	8	&	99	&	1	\\
IKT~7 	&	 CC	&	(d?) 	&	6	&	0	&	0	&	100	&	&	7	&	0	&	0	&	93	&	7	&	$\dotsm$	&	$\dotsm$	\\
B0050--728 	&	CC$^{\dagger}$	&	$\dotsm$	&	9	&	0	&	0	&	100	&	&	10	&	0	&	0	&	100	&	0	&	$\dotsm$	&	$\dotsm$	\\
IKT~16 	&	CC*	&	(d?)	&	6	&	0	&	0	&	100	&	&	6	&	0	&	0	&	93	&	7	&	$\dotsm$	&	$\dotsm$	\\
IKT~18 	&	CC$^{\dagger}$	&	$\dotsm$	&	62	&	91	&	0	&	9	&	&	63	&	92	&	0	&	8	&	0	&	$\dotsm$	&	$\dotsm$	\\
DEM~S108 	&	CC$^{\dagger}$	&	$\dotsm$	&	17	&	100	&	0	&	0	&	&	17	&	100	&	0	&	0	&	0	&	$\dotsm$	&	$\dotsm$	\\
IKT~21 	&	CC	&	$\dotsm$	&	139	&	92	&	0	&	8	&	&	141	&	93	&	0	&	7	&	0	&	87	&	13	\\
HFPK~334	&	Ia	&	$\dotsm$	&	3	&	0	&	0	&	0	&	&	3	&	0	&	0	&	0	&	100	&	96	&	4	\\
1E~0102.2--7219 	&	 CC*	&	(d?)	&	5	&	49	&	20	&	31	&	&	5	&	47	&	20	&	27	&	6	&	$\dotsm$	&	$\dotsm$	\\
IKT~23 	&	CC*	&	d	&	3	&	0	&	0	&	100	&	&	4	&	0	&	0	&	56	&	42	&	$\dotsm$	&	$\dotsm$	\\
DEM~S128	&	CC$^{\dagger}$	&	(d?)	&	25	&	76	&	15	&	9	&	&	26	&	75	&	16	&	8	&	1	&	$\dotsm$	&	$\dotsm$	\\
DEM~S130 	&	  CC$^{\dagger}$	&	(d?)	&	25	&	76	&	15	&	9	&	&	26	&	75	&	16	&	8	&	1	&	$\dotsm$	&	$\dotsm$	\\
IKT~25 	&	 CC*	&	(d?)	&	25	&	76	&	15	&	9	&	&	26	&	75	&	16	&	8	&	1	&	$\dotsm$	&	$\dotsm$	\\
N83C 	&	 CC$^{\dagger}$	&	$\dotsm$	&	32	&	100	&	0	&	0	&	&	32	&	99	&	0	&	0	&	1	&	$\dotsm$	&	$\dotsm$	\\
		\bottomrule
	\end{tabular}
}
	\begin{tablenotes}
		\small
		\item $N_{CC}/N_{Ia}$ represents the relative number of CC vs. Ia progenitors, and is derived using the Type Ia DTD of \citet{2017ApJ...848...25M} and the CC DTD assuming either a single or binary stellar population of \citet{2017arXiv170107032Z}.
		\item CC = Based on its SFH and its properties, this remnant arose from a CC SNe.
		\item Ia = Based on its SFH and its properties, this remnant likely arose from a Type Ia SNe.
		\item * = The SNR has independent evidence in supported of SN classification derived from the SFH.
		\item $^{\dagger}$ = Source associated with a H\textsc{II} region, dense environment, star forming region, stellar cluster or emission nebula.
		\item d = SFH is strongest in 50--100 Myr time bin, suggesting this SNR may have resulted from a delayed CC explosion.
		\item (d?) = SFH shows some SFH in the 50--100 Myr bin which may suggest a delayed CC explosion.
	\end{tablenotes}
\end{table*}

\subsection{Possible Type Ia SNRs}\label{iacc}

Four out of the 23 SMC SNRs have been suggested to arise from Type Ia SNe. This includes: IKT~5 [SNR 0047$-$73.5], IKT~25 [SNR 0104$-$72.3], IKT~4 [SNR J0048.4$-$7319], and DEM~S128 [SNR 0103$-$72.4]. As discussed in the previous section, IKT~5 and IKT~25 had SFHs and properties that are more consistent with CC SNRs than Type Ia SNRs. Thus, we focus on IKT~4 and DEM~S128 in this section.

Type Ia SNe occur in a wide range of environments, with little to intense star formation of both metal-poor and metal-rich populations \citep{2006A&A...453...57J, 2009ApJ...700..727B, 2013Sci...340..170W, 2014A&A...572A..38G, 2015MNRAS.448..732A}. To date, no SMC SNRs have confirmed classifications as Type Ia events, based on e.g., X-ray metal abundances or light echoes. Thus, our SFH analysis provides preliminary constraints on the nature of possible Type Ia SNRs in the SMC.

IKT~4 was suggested to be a Type Ia SNR based on the detection of enhanced Fe from a shallow {\it XMM-Newton} observation \citep{2004A&A...421.1031V}. The SF associated with IKT~4 exhibits an extended peak of intense metal-rich SF $<50$ Myr ago and a smaller peak of metal-poor SF around $\sim90$ Myr ago. Assuming that this remnant is from a Type Ia event, the probability that this source arose from a prompt progenitor is 98\%. This conclusion is similar to that found by \citet{2009ApJ...700..727B}, who showed that N103B (a SNR in an actively star-forming region in the LMC) likely arose from a young, prompt Type Ia progenitor.

The proximity of IKT~4 to the star-forming region N19 suggests it is possible that IKT~4 arose from a CC explosion rather than a Type Ia SN. The X-ray data of this source is not sufficiently deep enough to constrain its abundance properties, and thus the nature remains uncertain. If this remnant results from a CC of a massive star, our estimates from the SFH of this source indicate that 24\%,  17\% and 59\% of the CC progenitors have a mass of 8--12.5\,$M_{\odot}$, 12.5--21.5\,$M_{\odot}$  or $>21.5M_{\odot}$, respectively.

DEM~S128 was suggested to arise from a Type Ia SN based on centre-filled X-ray emission that exhibited enhanced Fe abundances \citep{2004A&A...421.1031V, 2015ApJ...803..106R}. However, the SFH of the stellar population associated with DEM~S128 shows intense, metal-rich SF that peaked $\sim$10~Myr ago, consistent with the SFH from a CC progenitor. Given that DEM~S128 is associated with a H\textsc{ii} region with the same name \citep{1995ApJS..101...41B}, as well as its recent, active SF, we believe DEM~S128 more likely resulted from a CC explosion rather than Type Ia event. Based on our estimates, 76\% of CC progenitors in the region are $>21.5M_{\odot}$ stars, and 9\% (15\%) are from 8--12.5\,$M_{\odot}$ (12.5--21.5\,$M_{\odot}$) stars, respectively. 

\subsection{Previously Unclassified SNRs}\label{uncla}

For nine out of the 23 SNRs in the SMC, their explosive origin is unknown or not well constrained. This includes DEM~S5 [SNR J0040.9--7337], B0045$-$733 [SNR J0047.5--7306], NS~21 [SNR J0047.8--7317], NS~19 [SNR J0048.1--7309], B0050--728 [SNR J0052.6--7238], DEM~S108 [SNR J0100.3--7134], HFPK~334 [SNR J0103.5--7247], DEM~S130 [SNR J0105.6--7204], N83C [SNR J0114.0--7317]. The main reason for their lack of classification comes from either the faintness of the source itself or due to shallow observations overlapping these sources. As such, combined with what information is available about each source, the SFHs of the stellar populations associated with these remnants can help us better constrain the possible SN origin. 

All nine remnants exhibit weak, thermal X-ray emission making it difficult to constrain the presence of enhanced abundances that can help constrain the nature of the SNe. However, optical observations of DEM~S5 detect bright [O \textsc{iii}] emission associated with its shell, which may suggest this remnant is an oxygen-rich CC SNR, much like 1E~0102.2--7219 and IKT~23, which also show bright [O \textsc{iii}] \citep{2007MNRAS.376.1793P}. However, some Type Ia SNRs (e.g, Kepler: \citealt{2007ApJ...662..998B, 2007ApJ...668L.135R, 2013ApJ...764...63B}; N103B: \citealt{2014ApJ...790..139W}) also exhibit strong  [O \textsc{iii}] emission and are found to be interacting with dense circumstellar material.
	
All of these previously unclassified remnants except for HFPK~334 are surrounded by stellar populations that exhibit strong SF $<50$ Myr ago. Their SFHs have well defined bursts of various intensities, consistent with that seen surrounding CC SNe. As nearly all of these remnants are found either to be interacting with dense material like DEM~S5 and N83C \citep{2008A&A...485...63F, 2015ApJ...799..158T}, or found close to a star forming region/cluster (e.g., B0045--733, NS~19, B0050-728, N83C), or H\textsc{II} regions (e.g., NS~21, B0050$-$728, DEM~S130), it is likely that these remnants arise from CC of massive stars. Under this assumption, the majority of the CC SN progenitors associated with these SNRs have a mass $>21.5M_{\odot}$. Of these, DEM~S5, NS~21, NS~19, DEM~S130 and N83C exhibit varying degrees of SF $\sim 50-200$ Myr ago, which may imply that a fraction of the stellar population underwent delayed CC. However, this fraction is no more than $\sim5$\% for most of these sources.  By contrast, the SF of DEM~S5 is dominated by significant SF $\sim$50--200 Myr ago, implying that $\sim56\%$ of the progenitors associated with this remnant underwent delayed CC. 

Compared to all other SNRs in the SMC, the SFH of HFPK~334 is quite unique. \citet{2014AJ....148...99C} showed that HFPK~334 is expanding into a low-density environment, a characteristic associated commonly with Type Ia SNRs, such as SN1006 \citep{1995Natur.378..255K}, RX J1713.7--3946 \citep{1997PASJ...49L...7K,1999ApJ...525..357S}, and SNR 0509--67.5 \citep{2004ApJ...608..261W}. HFPK~334 is the only SNR with virtually no SF in the last $\sim$50 Myr (although the uncertainties are large for the last $\sim$5 Myr). The peak SF of HFPK~334 occurred 50--1000 Myr ago, similar to the SFH observed for the Type Ia SNR N103B in the LMC \citep{2009ApJ...700..727B}. If HFPK~334 is the result of a CC explosion, then the lack of SF since 50~Myr ago indicates a delayed explosion due to rapid rotation or binary interaction. However, we are unable to use its SFH to place constraints on the possible mass of its progenitor, assuming it arose from a CC SN.
	
Given its low-density environment and its unique SFH, we suggest HFPK~334 likely arose from a Type Ia SN. Under this assumption, the SF between 64--180 Myr ago suggests it arose from a prompt Type Ia progenitor with a probability of 96\%. As such, unlike LMC SNR N103B, which showed a strong peak of emission $\sim12$ Myr suggesting this source may of have resulted from a young ($<150$Myr) progenitor, it is likely that HFPK~334 arose from an older ($>150$ Myr), $Z=0.004-0.008$ progenitor that underwent a prompt Type Ia explosion. As this remnant is expected to be relatively evolved (with an age of  $>1.4$ kyr), the X-ray spectrum is thought to be dominated by emission with ISM abundances \citep{2014AJ....148...99C}. Deeper observations would be required to confirm the nature of this source and determine whether its properties are consistent with that of Type Ia SN explosion models \citep[e.g.,][]{2008ApJ...680.1149B}.

\section{Discussion}\label{sect5}

\begin{figure*}[t]
	\begin{center}
		\includegraphics[width=\columnwidth]{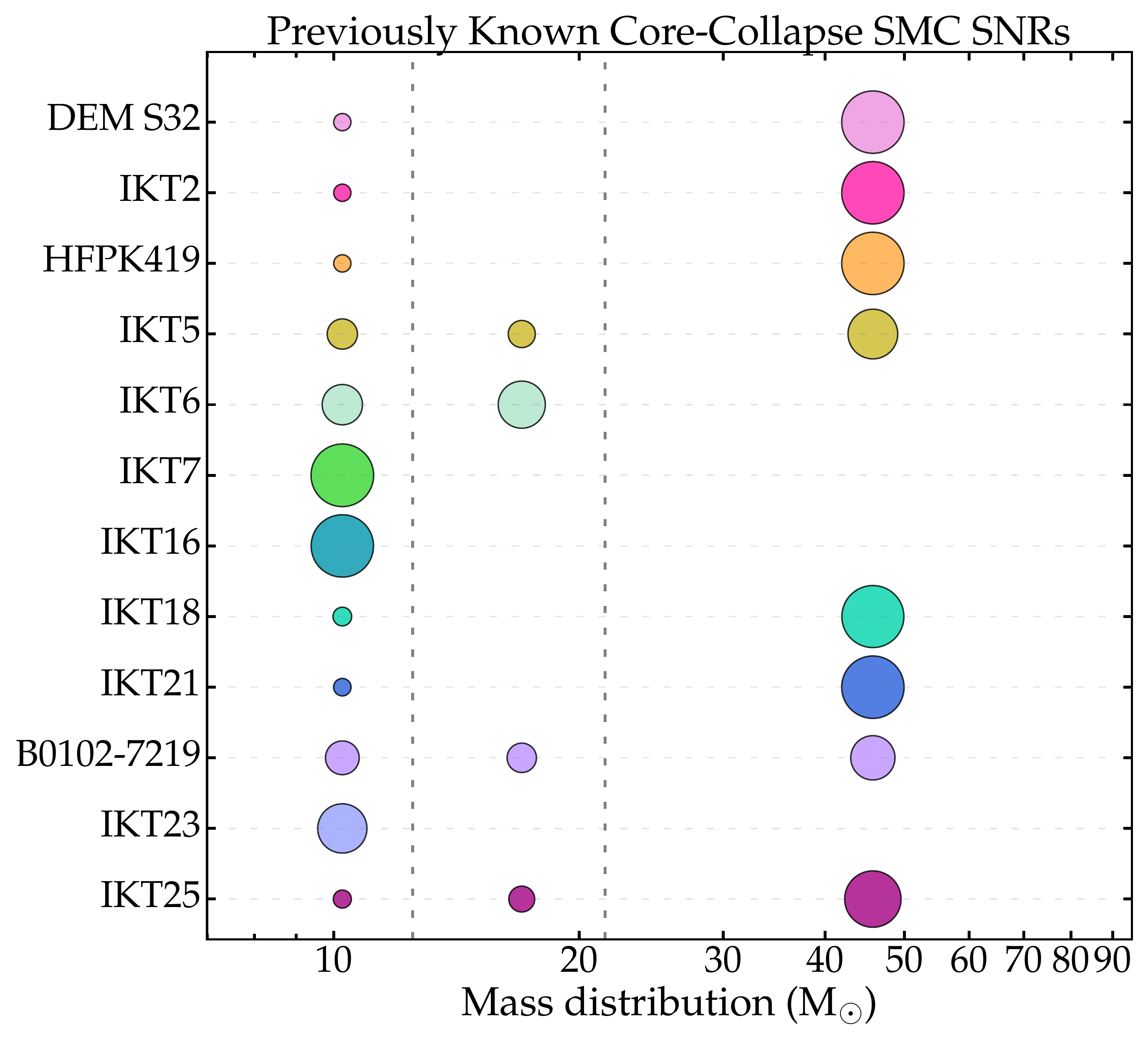}
		\includegraphics[width=\columnwidth]{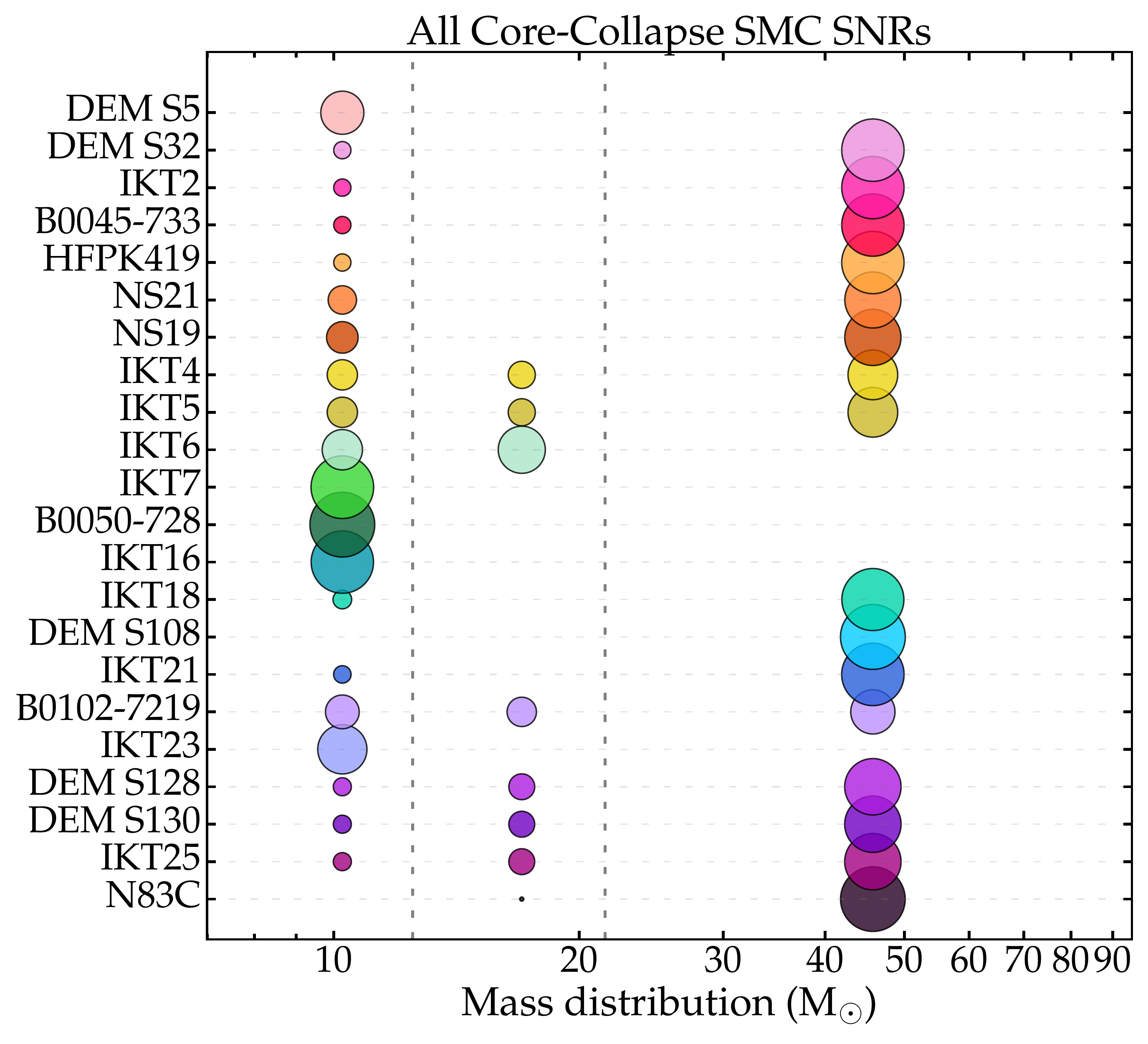}
		\caption{textit{Left:} Mass distribution of the previously classified CC SMC SNRs listed in Section \ref{cc}; \textit{Right:} all SNRs classified as CC in our analysis, assuming a CC DTD arising from a binary stellar population and a Salpeter IMF. The size of the plotmarker corresponds to how likely the SNR was formed by a progenitor of mass 8--12.5\,$M_{\odot}$, 12.5--21.5\,$M_{\odot}$ and $>21.5M_{\odot}$ assuming a binary star population. These plots visually represent the numbers given in Table \ref{CCSNRs}, with larger circles indicating greater likelihood, while no plot marker indicates a likelihood of zero. A similar trend is seen if one assumes a CC DTD and single-star population.}\label{massdist}
	\end{center}
\end{figure*}

We have examined the stellar populations in the vicinities of 23 known SNRs in the SMC. Using the SFH maps of \citet{2004AJ....127.1531H} and recent Type Ia and CC DTDs, we have attempted to characterise the nature of the progenitors of each SNR based on when the SFR peaked. We have compared our findings to the SNR properties, such as the metal abundances, the X-ray morphologies, and the surrounding environments. Despite the limitations of constraining the progenitors using the SFHs, we have found that the explosive origins ascertained from the SNR properties are consistent generally with those predicted from the proximal stellar populations. In particular, sources classified as CC SNRs from detailed spectral modelling and abundance estimates from deep X-ray studies had the best agreement with the SFH results (e.g., 1E~0102.2--7219: \citealt{2000ApJ...537..667B}, IKT~2: \citealt{2002PASJ...54...53Y}, IKT~23: \citealt{2003ApJ...598L..95P}, IKT~25: \citealt{2014ApJ...788....5L}).

\subsection{Comparison To Independent Progenitor Mass Estimates}

From our study, the vast majority of the stellar populations associated with the SNRs in the SMC have SFHs consistent with CC origins. Based on our DTD convolved local SFHs, we estimate the possible masses of the CC progenitors. In Figure \ref{massdist}, we have plotted the progenitor mass distribution of the known CC SNRs listed in Section \ref{cc} (left panel), and all SNRs classified as CC in our analysis (right panel). Here the area of the plot marker represents the likelihood that each source arose from a progenitor of mass 8--12.5\,$M_{\odot}$, 12.5--21.5\,$M_{\odot}$ or $>21.5M_{\odot}$, assuming a binary star population.  We find that most of the CC SMC SNRs have SFHs that are consistent with a progenitor mass $8-12.5M_{\odot}$ or $>21.5M_{\odot}$. 

For IKT~2, IKT~6, and IKT~25, which had estimates of the progenitor mass based on X-ray observations, our SFH mass estimates are consistent with these values. Previous work on 1E~0102.2$-$7219 suggested that the SNR had a 25--35$M_{\odot}$ progenitor \citep{2000ApJ...537..667B}, whereas our SFH method has difficulty distinguishing the progenitor mass due to multiple peaks in the SFR in the last 50~Myr. However, our results showed that 49\% of CC progenitors in the vicinity of 1E~0102.2$-$7219 are $>$21.5$M_{\odot}$, and 31\% of CC progenitors are 8--12.5\,$M_{\odot}$. Multi-wavelength studies by \cite{2003ApJ...598L..95P} indicated the progenitor of IKT~23 was a $\sim$18$M_{\odot}$ star, based on a comparison of the SNR's abundances (O/Ne, O/Mg, and O/Si) to the model predictions of \citet{1997NuPhA.616...79N}. However, the nearby stellar population suggest that 100\% of the CC progenitors arose from stars of 8--12.5\,$M_{\odot}$. We note that if one compares the SNR's abundance ratios to updated Type II~SN nucleosynthesis models from \citet{2016ApJ...821...38S}, then the ratios are consistent with a $\sim$15$M_{\odot}$ progenitor. Thus, in some cases, complementary optical and X-ray observations are crucial to differentiate the progenitor masses obtained from the SFH.

\subsection{The Initial Mass Function}

Based on various studies of SNe and SNRs in nearby galaxies (e.g., \citealt[][]{2014ApJ...791..105W, 2014ApJ...795..170J, 2015PASA...32...16S, 2018arXiv180207870D} and references therein), an absence of progenitors with mass $\gtrsim18M_{\odot}$ is apparent, suggesting a possible upper limit to the mass that produces observable CC SNe. Theoretically, this lack of progenitors above $\gtrsim18M_{\odot}$ could be a natural result of the fact that a fraction of massive stars are expected to implode and form a black hole without a visible SN \citep[e.g.,][]{2003ApJ...591..288H, 2016ApJ...821...38S}. This suggestion was supported by the discovery of the disappearing $\sim25M_{\odot}$ red supergiant in NGC~6946 that likely underwent a failed SN \citep{2015MNRAS.450.3289G, 2017MNRAS.468.4968A}. However, recent studies by \citet{2017MNRAS.469.2202M, 2018MNRAS.476.2629M} of the resolved stellar populations around Type IIP and stripped-envelope SNe, as well as the discovery of a possible progenitor system for Type Ic SN~2017ein \citep{2018MNRAS.480.2072K}, suggest that even progenitors $\gtrsim18M_{\odot}$ can produce observable explosions. However, due to a number of factors discussed in \citet[][]{2015PASA...32...16S}, the discovery rate of progenitor stars associated with SNe is small.

Nevertheless, combined with the properties of the SNR and its environment, the SFHs associated with these remnants provides a complementary and independent means to quantify the mass of CC progenitors. Based on our analysis, Figure \ref{massdist} suggests that a large fraction of the SMC's CC progenitors result from high mass progenitors, a result that contrasts the discoveries of pre-explosion SN progenitors \citep[e.g.,][]{2015PASA...32...16S} and studies of the SFH around SNRs in high metallicity galaxies M31/M33 \citep[e.g.,][]{2014ApJ...795..170J,2018arXiv180207870D}. 

\begin{figure*}[t]
	\begin{center}
		\includegraphics[width=0.99\columnwidth]{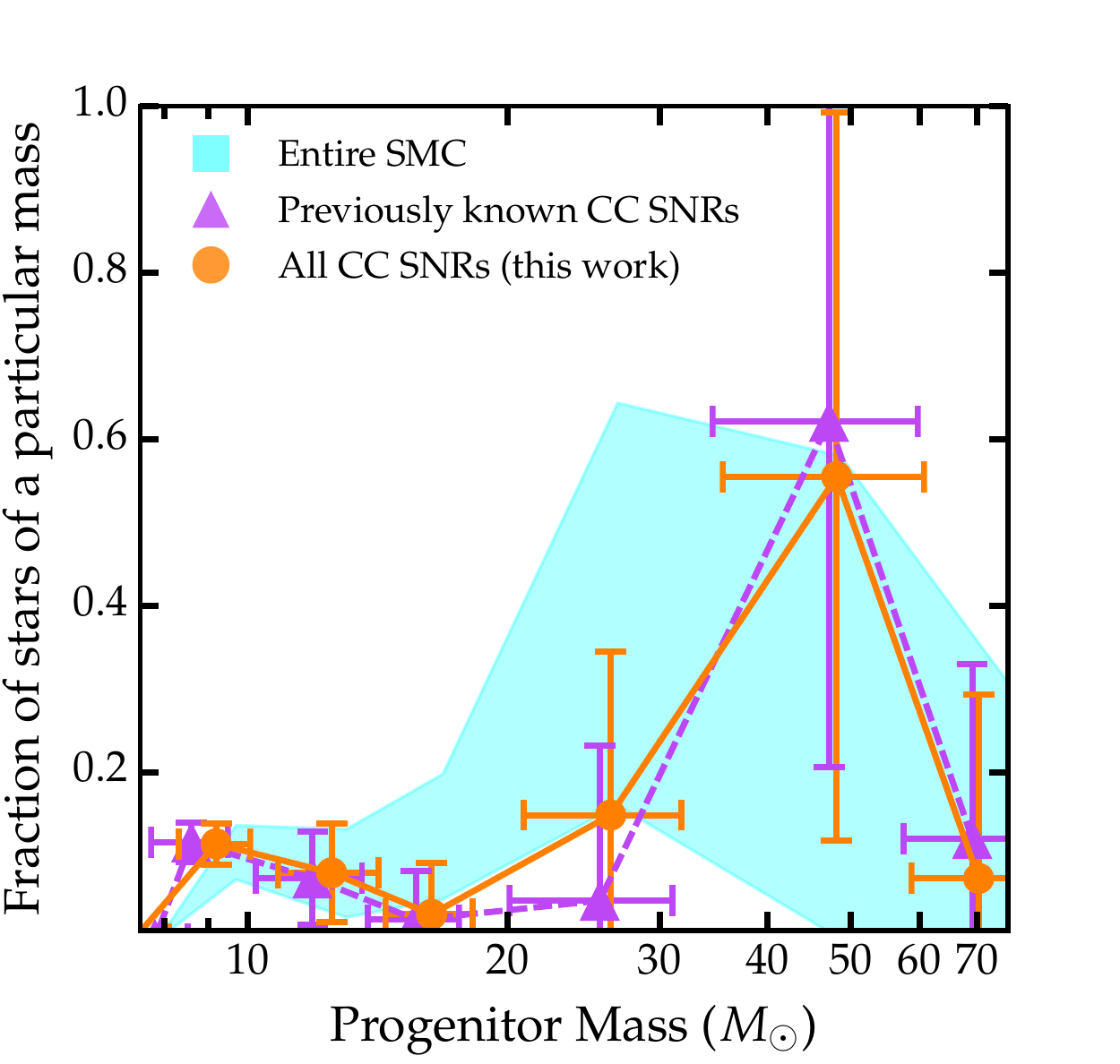}
		\includegraphics[width=0.98\columnwidth]{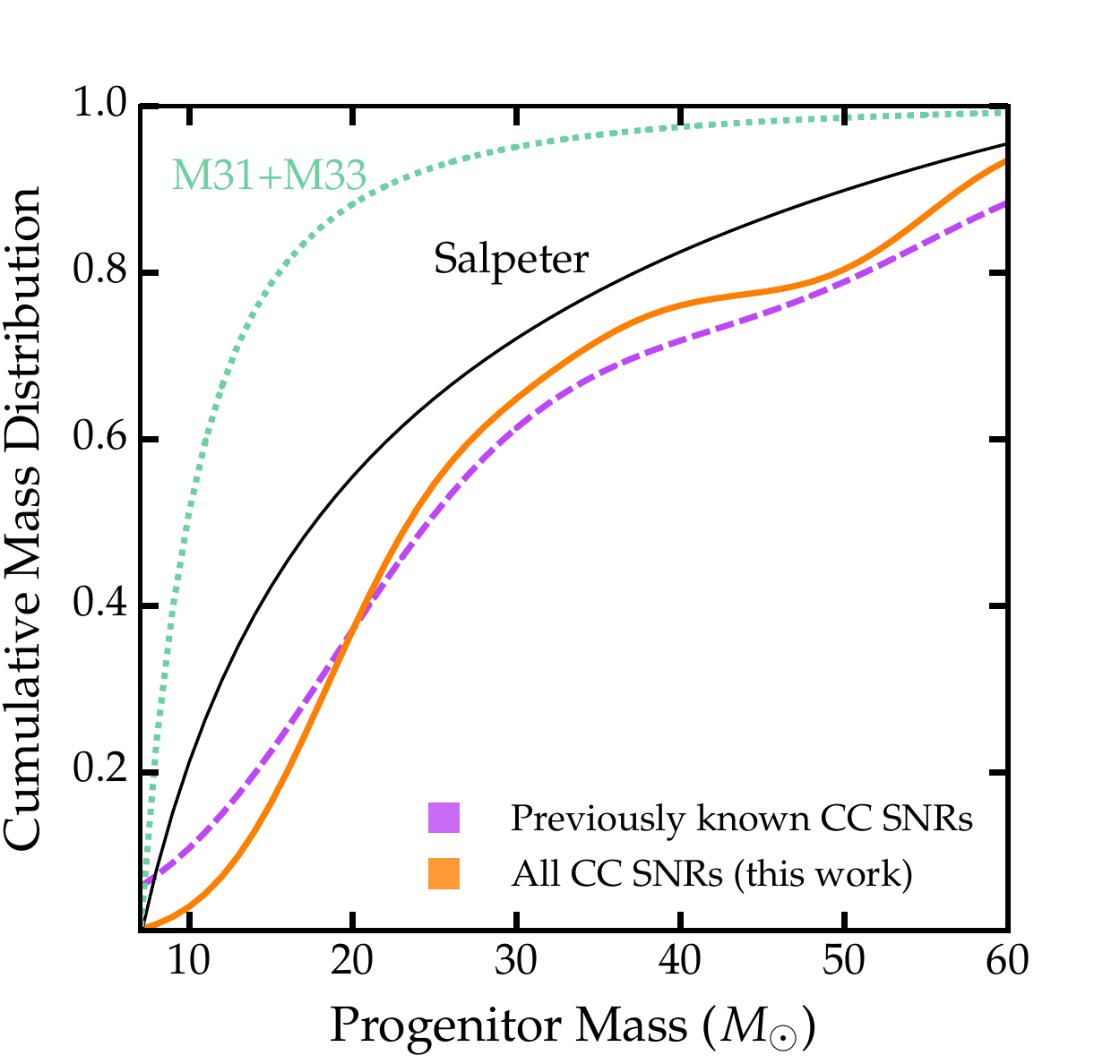}
		\caption{\textit{Left:} The fraction of SN progenitors of a particular mass associated with the previously known CC SNRs (purple triangles) and all SNRs classifed as CC in our analysis (orangle circle). To derive this, we convolved the stacked SFHs of each population with recent CC DTD from \citet{2017arXiv170107032Z}, and a Salpeter IMF. Here the error bars correspond to the 1$\sigma$ uncertainty in this fraction as derived from the uncertainties in the SFHs. The cyan shaded region corresponds to the 1$\sigma$ uncertainty in this fraction of SN progenitors of a particular mass assuming the average SFH of the entire SMC. One can see that the number of SN progenitors associated with the SNRs in the SMC is consistent with the fraction derived from the SFH of the entire SMC, implying that we are not overly biased in our analysis towards measuring more massive stars. \textit{Right:} The cumulative progenitor mass distribution derived using previously known CC SNRs listed in Table \ref{CCSNRs} (purple large dashed line), and all SNRs classified as CC in our analysis (orange dashed line). In addition, we include the equivalent distributions for a Salpeter IMF integrated to 120$M_{\odot}$ (black solid line), and the inferred mass distribution for M31 and M33 \citep[green dotted line;][]{2014ApJ...795..170J,2018arXiv180207870D}. One can see the mass distribution for the SMC is similar to that of a Salpeter IMF, but shallower than that seen in M31 and M33.} \label{cumulative}
	\end{center}
\end{figure*}

To probe this tension further, we investigate whether the SFHs at the sites of the SNRs yield a larger fraction of high-mass stars than would be expected from the SFH of the SMC as a whole. To do this, we sum the SFHs at the locations of the SNRs and convolve the stacked SFH with the recent CC DTD of \citet{2017arXiv170107032Z} and a Salpeter IMF. This method yields the number of stars formed as a function of lookback time, and we convert look-back time to progenitor mass using the single-star models of \citet{2017arXiv170107032Z}. This calculation then gives us the number of stars in each mass bin associated with the SNRs and with the SMC as a whole. Figure~\ref{massdist} (left panel) plots the fraction of stars in each mass bin (i.e., the number of stars in a mass bin divided by the total number of stars with masses between $\sim8-70~M_{\odot}$) for previously identified CC SNRs (purple triangles) and for all of the CC SNRs from this study (orange circles). The uncertainties in this fraction are derived from the uncertainty in the SFHs. Comparing the stacked SNR results to those of the SMC as a whole (the cyan band), we find that within the errors, these values are consistent. As such, our analysis is not biased toward identifying more massive progenitors than in the SMC as whole. 

In Figure \ref{cumulative} (right panel), we have plotted the cumulative progenitor mass distribution of previously known CC SNRs listed in Table \ref{CCSNRs} and of all SNRs classified as CC in our analysis as derived from their local SFHs. For reference, we include the corresponding cumulative distributions assuming a Salpeter IMF ($dN/dM \propto M^{-2.35}$) integrated to 120\,$M_{\odot}$, and the inferred mass distribution of the SNRs in M31 and M33 ($dN/dM \propto M^{-2.96}$: \citealt{2014ApJ...795..170J,2018arXiv180207870D}). 

We note that the mass distribution of the SNRs in the SMC is similar to that of a Salpeter IMF, but is not as steep as that seen in M31 and M33 which has fewer high-mass progenitors \citep{2014ApJ...795..170J,2018arXiv180207870D}. Assuming that the mass distribution of all the SMC SNRs can be well approximated using a power-law function of the form similar to that of the Salpeter IMF, we find that it can be well described using $dN/dM \propto M^{-1.84}$ (or $dN/dM \propto M^{-1.78}$ for the previously known CC SNRs).  This result implies that the SMC progenitor mass distribution is more top heavy than M31 and M33, with higher mass stars successfully producing a CC SNe.

Assuming that a Salpeter IMF is representative of the stellar distributions, our results suggest that SNRs, at least in the SMC, are associated with higher mass ($>21.5M_{\odot}$) progenitors. A similar conclusion is reached when one considers the progenitor mass estimates from the X-ray and optical properties of individual SNRs. For example, studies of the SMC SNRs IKT~25 and 1E~0102.2$-$7219 indicated progenitors with mass $>25M_{\odot}$ \citep{2014ApJ...788....5L, 2000ApJ...537..667B}, and the same tension is found with both LMC and Milky Way SNRs (e.g., G292.0+1.8: \citealt{2004ApJ...602L..33P}, G54.1+0.3: \citealt{2010ApJ...710..309T, 2015ApJ...807...30G}, W49B: \citealt{2013ApJ...764...50L}, MSH 11--61A: \citealt{2015ApJ...810...43A}, N49B: \citealt{2017ApJ...834..189P}, N132D: \citealt{2000ApJ...537..667B, 2018AAS...23124114P}). In contrast to these results, studies by e.g., \citet{2014ApJ...795..170J} and \cite{2018arXiv180207870D} found that the progenitor distribution of SNRs in M31 and M33, whose metallicity is higher than that of the SMC, is steeper than that of a Salpeter IMF based on the SFHs near the CC SNRs in those galaxies (see Figure \ref{cumulative} right). 

The fact that SNRs in the SMC are associated with more massive progenitors, suggesting that higher mass stars produce successful explosions, is also supported by recent theoretical work (e.g., \citealt{2003ApJ...591..288H, 2015ApJ...801...90P, 2014ApJ...783...10S, 2016ApJ...821...38S}). These authors showed that the outcome of CC SNe of a particular mass depends non-monotonically on the pre-supernova core structure of the massive star. By investigating the explosive outcomes of a wide variety of masses, these authors found that there is no single initial mass below which all stars explode and produce a neutron star or above which all stars would implode to a black hole. Rather, there are islands of explodability, in which even high-mass progenitors may explode and produce an observable remnant. As a consequence, SNRs may be expected to be associated with a wide variety of progenitor masses.

\subsection{The Effect of Metallicity on the IMF}

It has been suggested that low-metallicity environments tend to favour a more top-heavy IMF \citep[e.g.,][]{2002ApJ...571...30S, 2004ARA&A..42...79B, 2012MNRAS.422.2246M}. In these environments, the metal-poor molecular clouds are very efficient in forming massive stars since cooling via line emission or dust cooling as well as fragmentation is inefficient resulting in higher mass stars. \citep[e.g.,][]{2002ApJ...576..870P, 2004ARA&A..42...79B, 2005MNRAS.356.1201B, 2005MNRAS.359..211L,2012MNRAS.422.2246M}. However, even though understanding the effect of metallicity on the mass distribution of progenitors is fundamentally important for the theory of star formation, its current dependence is not well constrained in the literature \citep[see review by][]{2010ARA&A..48..339B}. However, there is increasing observational evidence which favours an IMF shallower than Salpeter in metal-poor environments such as those seen in ultra-faint dwarf \citep[e.g.,][]{2012ApJ...747...72D, 2013ApJ...771...29G} and early-type galaxies \citep{2015ApJ...806L..31M}. 

\subsection{The Effects of Mass Loss, Rapid Rotation and Binary Interaction on the IMF}

Apart from metallicity, both mass loss via stellar winds and binary interaction play an important role in altering the shape of the measured IMF \citep[e.g.,][]{2015ApJ...805...20S}. For single-star populations, two effects change the observed IMF of this population: stellar wind mass loss (which is dependent on the metallicity of the environment and the mass of the star) and when stars of a particular mass leave the main sequence (which translates into the slope of the IMF).  \citet{2015ApJ...805...20S} found that as stellar mass loss increases (which occurs with increasing mass and metallicity), the measured IMF of this population becomes more top-heavy.

For binary-star populations, two additional processes alter the observed mass function of these objects: stellar mergers \citep[e.g.,][]{1998MNRAS.298...93B} and mass exchange between binary components via Roche lobe overflow \citep[e.g.,][]{1967ZA.....65..251K}. Both of these processes tend to shift the stars towards higher masses \citep{2015ApJ...805...20S}, while also making the mass gainer (or merger product) appear younger since the fraction of burnt fuel decreases as fresh hydrogen is introduced in the core \citep[e.g.,][and reference therein]{2007MNRAS.376...61D}.  These rejuvenated binary products can make up nearly 1/3 of the stars $>40M_{\odot}$, but this fraction decreases as both the number of binary systems and the mass of the binary products decrease \citep{2015ApJ...805...20S}. As the amount of mixing will increase with greater mass, rotation rate and decreased metallicity \citep[e.g.,][]{2011A&A...530A.115B, 2013ApJ...764..166D, 2014ApJ...780..117S}, this effect will be more pronounced in low-metallicity environments like the SMC. Unresolved binaries also tend to make the observed mass function more top heavy \citep[e.g.,][]{2009MNRAS.393..663W, 2015ApJ...805...20S}, with this shift being strongest at larger stellar masses assuming a constant IMF slope. However, unresolved binaries tend to flatten the mass function less than resolved binaries. 

Assuming constant star formation, \citet{2014ApJ...782....7D} showed that $\sim$30\% of massive main-sequence stars are the products of binary interaction. Thus, they contribute a non-negligible fraction to the stellar populations measured to derive SFHs like that used in this study. However, most studies that calculate the SFH from a stellar population do not include interacting binaries, or they use an ad-hoc binary fraction in which secondary masses are drawn randomly from an IMF. However, work such as \citet{2017PASA...34...58E} has shown that the SFHs derived using single stars only tend to peak somewhat earlier, and not as strongly as the SFHs derive assuming binaries (see Figure 27 of \citealt{2017PASA...34...58E}). As such, ignoring interacting binaries can lead to incorrect SFHs being deduced.

In total, thirteen of the SMC SNRs have stellar populations that show evidence of bursts of SFR of various intensities $\sim$50--200 Myr ago (Figure \ref{SFH1}), and properties suggestive of CC origin. The additional burst of SF $\sim$50--200 Myr ago may hint towards evidence of delay CC SN timescales due to binary interaction. To quantify this, we use the CC DTD assuming the binary stellar population of \citet{2017arXiv170107032Z} to estimate the possible delay CC contribution. Here we assume that the $\sim$50--200 Myr peak results solely from massive progenitors undergoing CC, which allows us to estimate an upper limit of this contribution, which we list in Table \ref{CCSNRs}. We find that for most of the remnants, $<$8\% of the stellar population could come from a delayed CC as a result of binarity. However, for DEM~S5 and IKT~23, we find that a large fraction of the stellar population could have arisen from a delayed CC. As the properties of IKT~23 and DEM~S5 suggest that these source result from a CC event rather than a Type Ia SN, it is possible that these remnant arose from a delayed CC.

A natural consequence of binarity is that a large fraction of these massive stars will be rapidly rotating \citep{2013ApJ...764..166D}.  Detailed studies of stars in the SMC and LMC indicate that some stars are rotating much faster than those found in the Milky Way \citep[e.g.,][]{2007A&A...462..683M, 2014arXiv1411.3185G, 2017ApJ...846L...1D, 2017MNRAS.465.4795B}. \citet{2013ApJ...764..166D} suggested that $\sim20\%$ of massive main-sequence stars rotate rapidly due to binary interaction, while most, if not all, rapidly rotating stars are likely the result of mass transfer in these binaries. Under this assumption, they find that they can naturally explain the observed rotational distribution of stars in these samples. Rapid rotation increases chemical mixing within the interior of the star, which increases the fuel available to that star and naturally extend its lifetime \citep{1982ApJ...254..287S, 1998MNRAS.298..525P, 2000ARA&A..38..143M} to $\sim$50--200 Myr. 
	
We should also note that the SMC is known to be highly efficient in forming high-mass X-ray binaries (HMXBs) that consist of a neutron star orbiting a rapidly rotating B-type main-sequence star \citep[e.g.,][]{2005MNRAS.362..879S, 2016A&A...586A..81H}. This characteristic is consistent with evidence that binary interaction (and to some extent, low metallicity) favours the formation of rapidly-rotating stars \citep[e.g.,][]{2006MNRAS.370.2079D, 2010ApJ...716L.140A, 2015A&A...579A..44D, 2016A&A...586A..81H}.
	
\subsection{The Ratio of CC to Type Ia SN Progenitors}

To quantify the likelihood that each SNR in our sample arises from a CC or a Type Ia progenitor, we convolve the local SFH of each SNR with the Type Ia DTD of \citet{2017ApJ...848...25M} plus a CC DTD assuming either a single or binary stellar population from \citet{2017arXiv170107032Z}. To be consistent with the previous study of the resolved local SFH around SNRs in the LMC \citep[][]{2010MNRAS.407.1314M}, we divide the SFH of each SNR into three time bins: $<$35~Myr, 35--330 Myr, 330 Myr--14 Gyr. These bins represent the timescales of CC SNe, delayed CC SNe and prompt Type Ia SNe if one assumes binary star models, and delayed Type Ia SNe, respectively. To determine the ratio of expected CC to Ia progenitors ($N_{\rm CC}/N_{\rm Ia}$), we integrate the SFH in each bin to derive the mass formed at each timescale, and multiply this value by the Type Ia + CC SNe DTD. Here we have assumed that the distribution of stellar ages is representative of the stellar distribution when the original star underwent a SN. In Table \ref{CCSNRs}, we list the local $N_{\rm CC}/N_{\rm Ia}$ for each SNR assuming both a single-star or binary star population DTD for the CC SN.

Before convolving the Type Ia + CC SNe DTD with the SFH associated with each SNR, we convolved our DTD with the total $Z=0.008$ stellar-mass formed in the SMC (Figure \ref{totalSFH} left panel) to derive the ratio of CC to Ia progenitors for the SMC. We find $(N_{\rm CC}/N_{\rm Ia})_{\rm SMC}\sim6$ assuming both a single or binary stellar population. This ratio is similar to that derived from local SN surveys and in galaxy clusters. Using a volume limited SN sample, \citet{2011MNRAS.412.1441L} and  \citet{2017MNRAS.471.4966H} showed that $N_{\rm CC}/N_{\rm Ia}\sim3$; \citet{2007A&A...465..345D} and \citet{2007ApJ...667L..41S} estimated $N_{\rm CC}/N_{\rm Ia}\sim2-4$ based on elemental abundances in the intracluster medium from X-ray observations of galaxy clusters. Additionally, \citet{1995MNRAS.277..945T} derived $N_{\rm CC}/N_{\rm Ia}\sim3-6$ for the solar neighbourhood and the Magellanic Clouds using the observed elemental abundances and comparing them to galactic models of chemical evolution. More recently, \citet{2016A&A...585A.162M} estimated $N_{\rm CC}/N_{\rm Ia}\sim1.2-1.8$ for the LMC, assuming that the number of observed LMC SNRs is representative of the CC/Ia SN rates.

\citet{2016A&A...585A.162M} suggested that the discrepancy between the LMC $N_{\rm CC}/N_{\rm Ia}$ ratio to those derived from local SN surveys and the intracluster medium is a result of the unique SFH of the Magellanic Clouds (which have had several distinct epochs of active star formation). However, in their original study \citet{2016A&A...585A.162M} did not utilise a DTD in their analysis when attempting to interpret the local SFH of each LMC remnant in terms of a SN progenitor. As such, it is possible that their $N_{\rm CC}/N_{\rm Ia}$ ratio is underestimated and in reality the $N_{\rm CC}/N_{\rm Ia}$ ratio for the LMC is consistent with the observational studies listed above.

For each SNR in the SMC, their SFH suggest $N_{\rm CC}/N_{\rm Ia}>1$, implying that these remnants are associated with stellar populations that favour a CC progenitor over that of a Type Ia. Among the sources which have  $N_{\rm CC}/N_{\rm Ia}<(N_{\rm CC}/N_{\rm Ia})_{\rm SMC}$, where $(N_{\rm CC}/N_{\rm Ia})_{\rm SMC}\sim 6$,  IKT~6 has an associated neutron star, IKT~23 and 1E~0102.2--7219 are oxygen-rich CC SNRs, and DEM~S5 and NS~21 are located near H\textsc{ii} regions. Consequently, the properties of these SNRs are more consistent with CC SNe than with Type Ia SNe. HFPK~334 has one of the lowest ratios ($N_{\rm CC}/N_{\rm Ia}\sim3$), implying a preference of CC over that of Type Ia progenitors. Based on its low-density environment (see Section \ref{uncla}), and its SFH which is distinct from all other SNRs in our sample,  we believe HFPK~334 likely arose from a Type Ia SN. Another possibility is that this SNR arose from a delayed CC SN.

\subsection{Selection Effects}

One remarkable result from our analysis is that nearly all (22/23) remnants in the SMC have SFHs and properties consistent with CC events. The four SNRs (IKT~4 [SNR J0048.4$-$7319]; IKT~5 [SNR 0047$-$73.5]; IKT~25 [SNR 0104$-$72.3]; DEM~S128 [SNR 0103$-$72.4]) suggested to arise from Type Ia SNe in the literature all have properties, local SFHs, and $N_{\rm CC}/N_{\rm Ia}$ indicative of CC explosions. However, it is noteworthy that $(N_{\rm CC}/N_{\rm Ia})_{\rm SMC}$ is lower than that implied by our classifications of the SNR sample ($N_{\rm CC}/N_{\rm Ia}=22/1$). For the SMC SNRs to have $(N_{\rm CC}/N_{\rm Ia})_{\rm SMC}\sim6$, it should have $\sim$3--4 Type Ia SNRs (out of 23 known SNRs), which is lower than that suggested from our analysis. 

The discrepancy between the high $N_{\rm CC}/N_{\rm Ia}$ of the SMC implied by Table \ref{CCSNRs} and the measured $(N_{\rm CC}/N_{\rm Ia})_{\rm SMC}\sim6$ arises from the scarcity of identified Type Ia SNRs in the SMC. \citet{1998ApJ...503L.155K} and \citet{2000A&A...362.1046L} suggested that the rates of Type Ia SNe are metallicity dependent, with low-metallicity environments inhibiting Type Ia SNe because the wind of the accreting white dwarf is too weak to reach the Chandraskhar mass limit. However, Type Ia SNe are observed in low-metallicity galaxies (e.g., \citealt{2008ApJ...673..999P}), and several studies have argued that the Type Ia rate should {\it increase} as metallicity decreases because low-metallicity stars produce higher-mass white dwarfs \citep{1999ApJ...513..861U, 2008A&A...487..625M, 2013ApJ...770...88K}. 

Recent work by \citet{2017MNRAS.464.2326S} suggested that Type Ia SNRs may be more difficult to detect than CC SNRs in the SMC. These authors showed through semi-analytical modeling that a large fraction ($\sim30-40$\%) of Local Group SNRs are missed in current radio surveys. They found that most SNRs above detection limits will be CC SNRs, whereas Type Ia SNRs evolve in lower ambient densities and have lower surface brightnesses and thus shorter visibility times, which may preclude their detection. 

This result naturally arises from the fact that the luminosity of a source is proportional to the square of the density $n$ of the surrounding environment \citep[e.g.,][]{2015ApJ...803..101P}. As the massive progenitors of CC SNe tend not to travel far from their original birth sites due to their short lifetimes \citep{2011MNRAS.414.3501E, 2018arXiv180409164R}, these sources are usually found in dense environments. As a result, these SNRs tend to be more X-ray bright than those found in low-density environments \citep[e.g.,][]{2015ApJ...803..101P}. 

In addition, the prevalence of CC SNRs in the SMC may be a natural result of the mass-loss properties of the progenitors implied by the SFHs. As the mass-loss rate (\.{M}) increases, the X-ray luminosity of stellar ejecta and swept-up material will scale as \.{M}$^{2}$ \citep[e.g.,][]{2015ApJ...803..101P}. As higher-mass progenitors tend to lose more mass prior to explosion, the X-ray luminosity of those SNRs will be significantly brighter than those resulting from lower-mass progenitors. Thus, these SNRs will be detected more readily, introducing a selection effect that SNRs detected in X-rays arise from higher-mass CC SN progenitors.
	
However, we note that $n$ naturally influences the observable lifetimes of SNRs. SNR evolution is described as three distinct stages: first is the free expansion phase, where the shock front travels unimpeded by the surrounding environment \citep{1982ApJ...258..790C, 1999ApJS..120..299T}. Subsequently, the Sedov-Taylor phase \citep{1950RSPSA.201..159T,sedov59} occurs when the mass of swept-up material is comparable to the mass of ejecta, causing the forward shock to decelerate and producing a reverse shock that heats the inner ejecta to X-ray emitting temperatures \citep{1974ApJ...188..335M}.  This phase ends when radiative cooling becomes dynamically important, and the remnant enters the snow-plough phase \citep{1977ApJ...218..148M} before disappearing. The transition between the Sedov-Taylor phase and the radiative snow-plough stage occurs at $t = 2.9\times10^{4} E_{51}^{4/17} n^{-9/17}~{\rm years}$ (where $E_{51} \equiv E/10^{51}$~erg is the explosion energy; \citealt{1998ApJ...500..342B}) and can be used as an approximate lifetime of the SNR. Assuming $E_{51} = 1$, the lifetime of a SNR in a dense environment will be shorter than one in a low-density environment. As a result, CC SNRs arising from more massive progenitors are likely to be brighter than those from lower-mass progenitors but tend to have much shorter lifetimes.

Since the original \textit{XMM-Newton} X-ray \citep{haberl12} and ATCA radio \citep{2004MNRAS.355...44P, 2005MNRAS.364..217F} surveys of the SMC are quite shallow\footnote{The \textit{XMM-Newton} survey covered the galaxy with an effective exposure of $\sim$25~ks per field, corresponding to a flux limit of $\sim10^{-14}$ erg~s~cm$^{-1}$ \citep{haberl12}. The ATCA survey of the SMC observed SNRs down to a flux of $\sim2\times10^{16}$ W Hz$^{-1}$ at 1.42 GHz \citep{2004MNRAS.355...44P, 2005MNRAS.364..217F}.}, it is possible that we are currently biased towards detected the remnants from the highest mass progenitors, while some fraction of SNRs, are below detection limits.  In fact, it was shown by e.g., \citet[][]{2009ApJ...703..370C} that current radio SNR surveys are sensitivity-limited. Thus, our results may suggest that the SNR sample in the SMC is incomplete, and a deeper, systematic study of the SMC at multiple wavelengths would be beneficial to identify and characterise the full SNR population.

\section{Conclusion}\label{sect6}

In this paper, we present a systematic study of the stellar populations in the vicinities of 23 known SNRs in the SMC. Combined with the properties of the remnant themselves and their surrounding environment, we investigate the SFH at the sites of the SNRs, and we infer the natures of the SN progenitors based on the time since peaks in the SFR. The explosive origins of many SMC SNRs have not been characterised previously, and the SFHs reveal the likelihood of whether individual sources arise from a CC or Type Ia SN. We find that nearly all known SNRs (22/23) have local SFHs and properties consistent with CC SNe, including four remnants that had previously been classified as Type Ia SNRs in the literature based on their X-ray properties. The scarcity of Type Ia SNRs may be because these sources are visible for much shorter times than CC SNRs, and are intrinsically less luminous \citep{2017MNRAS.464.2326S}, or that CC remnants tend to be easier to detect due to the nature of their progenitor.

By convolving the local SFHs of each remnant with recent single and binary stellar population CC DTDs, we estimate the mass distribution of CC progenitors in the SMC. We find that this distribution is consistent with a standard Salpeter IMF, but shallower than that measured in M31 and M33 using similar methods, implying that SNRs in the SMC are associated with more massive stars. This is consistent with individual X-ray and optical studies of particular SMC SNRs (e.g., IKT~25 and 1E~0102.2$-$7219) which had suggested large progenitor masses ($\sim$25$M_{\odot}$) for these remnants previously. The top-heavy mass distribution suggested by these works contrasts the progenitor masses of $\ls$18$M_{\odot}$ estimated from stellar population studies around SNRs and pre-explosion images of SNe in nearby galaxies. However, recent theoretical work by e.g., \citet[][]{2014ApJ...783...10S, 2015ApJ...801...90P, 2016ApJ...821...38S} have shown that there is no single mass below or above in which a progenitor will explode or implode, with even progenitor masses $\gs$18$M_{\odot}$ able to produce observable explosions.

Furthermore, we show that a large fraction of the SMC SNRs exhibited a burst of SF between 50--200~Myr ago.  For example, the oxygen-rich SNR IKT~23 likely had a massive progenitor, and its peak SF occurred $\sim$50--200~Myr ago. As such, it is possible that IKT~23 and other SMC SNRs had progenitor stars with extended lifetimes, either due to their environments or to binarity. As most of these sources have properties and SFHs consistent with CC origins, we suggest that the long massive-star lifetimes may be a product of binary interaction and rapid rotation \citep[e.g.,][]{2011A&A...530A.115B, 2013ApJ...764..166D, 2015ApJ...805...20S, 2017arXiv170107032Z}, or the low-metallicity environment of the SMC \citep[e.g.,][]{1982ApJ...254..287S, 1998MNRAS.298..525P, 2000ARA&A..38..143M}.

\bigskip

\textit{Acknowledgements:} We thank the referee for their very detailed and insightful review that helped improve both the quality and clarity of the work presented. We thank Nicole Man, Marc Pinsonneault and Jeremiah Murphy for their helpful discussions. LAL gratefully acknowledges that this work was supported through NSF Astronomy \& Astrophysics Grant AST$-$1517021, and acknowledges support from the Sophie and Tycho
Brahe Visiting Professorship at the Niels Bohr Institute. CB is supported by grants NSF/AST-1412980 and NASA ADAP NNX15AM03G-S01. ERR is supported by the David and Lucile Packard Foundation and the Danish National Research Foundation through a Niels Bohr Professorship.  JFB is supported by NSF grant PHY-1714479.

\begin{appendix}\label{app}
	
\section{Properties of the supernova Remnants in the SMC}

\textbf{DEM~S5} (\textit{B0039$-$7353, HFPK~530, SNR~J0040.9$-$7337, SNR~J004100$-$733648)}: One of the largest SNRs (diameter $\sim$60 pc) in the SMC, DEM~S5 was first classified as a SNR based on the detection of X-ray emission from the position of an emission nebula from \emph{ROSAT} \citep{1999A&AS..136...81K, 2000A&AS..142...41H}. Its SNR origin was later verified by \citet{2005MNRAS.364..217F} and \citet{2007MNRAS.376.1793P} using ATCA. This source has a complex optical shell comprised of two intersecting shells, and bright [O \textsc{iii}] emission suggests that the SNR shock front is travelling with $v>100$ km/s. Shallow \textit{XMM-Newton} observations of this source detected faint X-ray emission coincident with bright H$\alpha$ and [S \textsc{II}] emission, implying that the SNR is interacting with dense material \citep{2008A&A...485...63F}.  Due to the shallowness of the \textit{XMM-Newton} observation and low surface brightness of the source, no abundance measurements have been constrained, although its size, low temperature, and ionisation timescale indicate an old age \citep{2008A&A...485...63F}. As of this writing, no attempt has been made to characterise the progenitor type of this remnant.

The SFH of DEM~S5 indicates that the SNR is located in an active star-forming region of the SMC. The SFH is dominated by metal-rich SF $\sim$50 Myr ago, with a smaller peak $\sim$30 Myr ago. A low-metallicity SF event occurred $\sim$100~Myr ago. Assuming that the SNR arose from a CC SN, 100\% of the progenitors have masses of 8--12\,$M_{\odot}$. We note that significant SF happened $\sim$50--200 Myr ago, which may imply that a large fraction (56\%) of the progenitors associated with this remnant underwent delayed CC. If DEM~S5 arise from a Type Ia SN, then a prompt progenitor is favoured for this source given the low SFR at lookback times $>$100 Myr. Follow-up observations of DEM~S5 are necessary to confirm the nature of the progenitor. 

\textbf{DEM~S32} (\textit{SNR~J0046.6$-$7309}): After this source was detected in a number of X-ray surveys of the SMC, \citet{2004A&A...421.1031V} obtained a pointed \textit{XMM-Newton} observation of DEM~S32. They found significant thermal X-ray emission, with evidence of emission lines from O, Ne, Si, Fe, and possibly Mg. Assuming a Sedov model of SNR evolution \citep{1959sdmm.book.....S}, they estimated that the SNR is $\sim6$ kyr old and has swept up $\sim$43 M$_{\odot}$ of material. Due to its location close to large nebula N19 \citep{1956ApJS....2..315H,2001AJ....122..849D}, \citet{2004A&A...421.1031V} suggested that the SNR was a CC explosion. This source is located in the same SF subcell as HFPK~419 and IKT~2. As a consequence, this source has the same SFH reported here as these SNRs. According to our estimates from the SFH, this SNR most likely arose from a $>21.5M_{\odot}$ progenitor that underwent a Type IIP explosion, similar to HFPK~419.

\textbf{IKT~2} (\textit{SNR~J0047.2$-$7308, HFPK~413, B0045$-$73.4}): Detected across radio wavelengths \citep[e.g.,][]{2001AJ....122..849D, 2004MNRAS.355...44P, 2009AJ....138.1101L}, this source was first studied in X-rays by \citet{2002PASJ...54...53Y} using \emph{ASCA}. They found that the X-ray morphology is center-filled and highly irregular, and the X-ray emission has enhanced abundances of Ne and Mg consistent with a progenitor of $\sim20M_{\odot}$. Using \emph{ROSAT} and \emph{XMM-Newton}, \citet{2001AJ....122..849D} and \citet{2004A&A...421.1031V} found that the X-ray emission from IKT~2 is best described by a thermal plasma model with enhanced Ne, Mg, Si and possibly Fe, suggesting an ejecta origin for the emission. \citet{2004A&A...421.1031V} derived a Sedov age for the SNR of $\sim5.6$ kyr, and they classified the SNR as from a CC explosion based on enhancement of intermediate-mass elements and location near the nebula N19. By comparing their abundance measurements to SN explosion models, \citet{2002PASJ...54...53Y} concluded that the SNR most likely arose from a $\sim20M_{\odot}$ progenitor.  Using \emph{Chandra}, \citet{2006AAS...20915614W} detected hard X-ray emission within the remnant, possibly from a  PWN. It was originally suggested that both IKT~2 and HFPK~419 were one remnant due to their proximity, but \citet{2004A&A...421.1031V} showed based on their X-ray properties that these sources are distinct. Found in the same subcell as HFPK~419 and DEM~S32, this SNR exhibits the same SFH as the two other remnants. This SNR is located within the nebula N~19, which is associated with moderate starburst activity \citep{2001AJ....122..849D}. From the SFH, 92\% of the CC SN progenitors in this region are from stars $>21.5M_{\odot}$. This value is slightly greater than the progenitor mass of 20$M_{\odot}$ suggested by \citet{2002PASJ...54...53Y}, but it is consistent within their errors. If the pulsar wind nebula identified by \citet{2006AAS...20915614W} is confirmed, then this result is consistent with the models of \citet{2016ApJ...821...38S} which suggested that neutron stars are produced in SNe of stars $<30M_{\odot}$ .

\textbf{B0045$-$733} (\textit{HFPK~401, SNR~J0047.5$-$7306}): This compact ($\sim$13 pc in diameter) SNR was first detected in X-rays by \citet{haberl12} using \emph{XMM-Newton}. Due to its low surface brightness, there is little known about its properties; however, \citet{2005MNRAS.364..217F} and \citet{2007MNRAS.376.1793P} detected [O \textsc{ii}] and H$\beta$ emission from the source, confirming its SNR nature. The SNR is located near star-forming nebula N19 and is found in the same subcell as HFPK~419, DEM~S32, and IKT~2. As such, B0045$-$733 has the same SFH history as reported for these SNRs, with significant metal-rich SF around $\sim$5 Myr ago and $\sim$30~Myr ago, consistent with a CC origin. Assuming a CC event, 92\% of progenitors in this region arose from stars $>21.5M_{\odot}$.

\textbf{HFPK~419} (\textit{SNR~J0047.7$-$7310}): Overlapping SNR IKT~2, this source is located within the emission nebula N19, which is thought to be associated with modest starburst activity \citep{2001AJ....122..849D}. The SNR is located in a part of N19 which had a period of intense, metal rich ($Z=0.008$) star formation $\sim$5 Myr ago as well as another burst $\sim$30 Myr ago (see Figure \ref{SFHfull1} and \ref{SFH1}). The region has had minimal SF activity aside from these events. \citet{2004A&A...421.1031V} suggested that the SNR originated from a CC SN based on detection of enhanced Ne and Mg relative to Si and Fe, but they did not constrain a progenitor mass nor whether the X-ray emission arose from ejecta or from ISM material. The asymmetric X-ray morphology of the SNR \citep[see Figure 2 of][]{2004A&A...421.1031V} is also consistent with a CC event \citep[e.g.,][]{2009ApJ...706L.106L, 2009ApJ...691..875L, 2011ApJ...732..114L}. \citet{2004A&A...421.1031V} estimated a Sedov age of $\sim8.5$ kyr for this remnant.  Based on the SFH there (assuming a single star population), it is expected that the majority (92\%) of the CC SN progenitors in the sub-cell have a mass $>21.5M_{\odot}$, with the other 8\% having masses of 8--12.5\,$M_{\odot}$. As per the classifications defined in Section~\ref{progenitors}, this result suggests the SNR arose from either a Type IIP or a Type Ib/Ic SN. However, as only a shallow {\it XMM-Newton} observation is available of HFPK~419 \citep{2004A&A...421.1031V}, further study is warranted to confirm the nature of this source. 

\textbf{NS~21} (\textit{DEM~S35, SNR~J0047.8$-$7317}): This radio-detected SNR is associated with the H\textsc{ii} region DEM S35 \citep{2005MNRAS.364..217F,2012ApJ...755...40P}. \citet{haberl12} detected no X-rays arising from the position of NS~21 in their \emph{XMM-Newton} survey of the SMC. Studying the lifetime and destruction efficiencies of silicate and carbon dust in the Magellanic Clouds, \citet{2015ApJ...799..158T} found that the dust-to-gas ratio surrounding this source is enhanced, implying a dense environment. NS~21 has a similar SFH as IKT~4 and IKT~5, which are located nearby. The SFH has a peak of metal-rich SF $\sim$9 Myr ago and extended, metal-rich SF from $\sim$30--1000 Myr ago. Assuming the SNR arises from a CC SN, 22\% of CC progenitors would be 8--12.5\,$M_{\odot}$ stars, and 78\% would be stars $>21.5M_{\odot}$. Assuming a Salpeter IMF, 34\% of stars would have $>40M_{\odot}$. NS~21 exhibits a strong peak of SF in the 50--200 Myr bin, which could be a signature of delayed CC. However, if NS~21 was a Type Ia SN, then we find that 99\% of progenitors are prompt, and 1\% are delayed. Deep X-ray observations of NS~21 to measure the plasma abundances would be beneficial to constrain the nature of the SNR.

\textbf{NS~19} (\textit{DEM~S31, SNR~J0048.1$-$7309}): No X-ray emission has been detected that is coincident with the optical emission from NS~19. However, \citet{haberl12} detected a larger, elliptically-shaped source of X-ray emission which may be related to the SNR. The radio emission of the source arises from the southwest portion of the remnant \citep{2005MNRAS.364..217F}; however, contamination from a nearby H\textsc{ii} region \citep{2001AJ....122..849D} makes it difficult to disentangle the SNR properties. It is located near the SNRs HFPK~419, DEM~S32, and IKT~2 and the N19 star-forming region. The SFH of NS~19 is similar to those of the nearby remnants, with SF peaks at $\sim$5~Myr ago and $\sim$30~Myr ago. Based on our estimates, 26\% of the CC progenitors are stars with masses between 8--12\,$M_{\odot}$, while 74\% arise from stars with masses $>21.5M_{\odot}$. Using a Salpeter IMF, 34\% of $>21.5M_{\odot}$ stars will be $>40M_{\odot}$.

\textbf{IKT~4} (\textit{HFPK~454, SNR~J0048.4$-$7319}): Using \emph{XMM-Newton}, \citet{2004A&A...421.1031V} found that the X-ray emission from IKT~4 peaks in the 0.7--1.0 keV band, implying that its emission is dominated by Fe-L emission. \citet{2004A&A...421.1031V} showed the X-rays arise from a smaller radius than the optical emission of this source \citep{1984ApJS...55..189M}, suggesting it is an old SNR. These authors were unable to estimate the elemental abundances because the SNR is faint in X-rays, but the Fe-L emission may indicate a Type Ia explosive origin. As IKT~4 is found in the same subcell as IKT~5, it has the same SFH, with extended but intense SF $<$50 Myrs ago. More details of its SFH can be found in Section \ref{iacc}.

\textbf{IKT~5} (\textit{DEM~S49, HFPK~437, B0047$-$735, SNR~0047$-$73.5, SNR~J0049.1$-$7314}): This radio-dim SNR \citep{1984ApJS...55..189M} was first discovered in soft X-rays using the \textit{Einstein Observatory} \citep{1983IAUS..101..535I}. Deep, follow-up observations using \emph{XMM-Newton} and \emph{Chandra} by \citet{2004A&A...421.1031V} and \citet{2015ApJ...803..106R}, respectively, found extended X-ray emission which is best described by a thermal plasma model with possibly enhanced Ne, Mg, and Fe. They detected a point source coincident with IKT~5 but claimed it was most likely a star. Using ATCA, \citet{2015ApJ...803..106R} also found a radio half-shell coincident with the X-ray emission, which is connected via an optical shell \citep{2007MNRAS.376.1793P}. Both \citet{2004A&A...421.1031V} and \citet{2015ApJ...803..106R} suggested a Type Ia origin based on the enhanced abundances. However, \citet{2015ApJ...803..106R} also suggested that this source may be from a CC given its environment and enhanced Ne and Mg abundances. \citet{2005MNRAS.362..879S} identified a hard X-ray point source at the remnant's center that is a high-mass X-ray binary (HMXB) candidate located near the SNR's geometric center \citep[][Auchettl et al. 2018, in prep]{2005MNRAS.362..879S}.  The vicinity of IKT~5 shows consistent, strong metal-rich SF for the last $\sim$50 Myr with a peak SFR $\sim$30 Myr ago. A burst of metal-poor SF occurred $\sim$100 Myr ago. We find that 24\% of CC progenitors are stars with masses of 8--12.5\,$M_{\odot}$, 17\% have a mass 12.5--21.5\,$M_{\odot}$, while 59\% have masses $>$21$M_{\odot}$.  Assuming a binary stellar population DTD, the enhanced SF seen between 50--100 Myrs (Figure \ref{SFH1}) suggests that 2\% of the progenitors underwent delayed CC due to binarity. 

\textbf{IKT~6} (\textit{1E~0049.4$-$7339, HFPK~461, B0049$-$73.6, SNR~J0051.1$-$7321}): IKT~6 was first discovered in X-ray surveys \citep{1983IAUS..101..535I, 1992ApJS...78..391W, 2000A&AS..142...41H} and followed up with \emph{XMM-Newton} \citep{2004A&A...421.1031V}, \emph{Chandra} \citep{2005ApJ...622L.117H, 2014ApJ...791...50S}, and \emph{Suzaku} \citep{2016PASJ...68S...9T}. IKT~6 exhibits a centre-filled X-ray morphology, with a distinct ring-like feature in projection that is surrounded by a larger, fainter, metal-poor shell. The SNR has enhanced O, Ne, Mg and Si abundances in its center, indicating that the X-ray emission arises from ejecta. \citet{2004A&A...421.1031V}, \citet{2005ApJ...622L.117H}, and \citet{2014ApJ...791...50S} determined that the SNR is evolved, with a Sedov age of $\sim14-17$kyr. By comparing the properties of IKT~6 to other Type II SNRs in the SMC (e.g., IKT~22), \citet{2004A&A...421.1031V} suggested that the SNR was from a CC explosion. More recently, \citet{2005ApJ...622L.117H} and \citet{2014ApJ...791...50S} suggested that this SNR resulted from an asymmetric CC SN of a 13--15$M_{\odot}$ progenitor with solar or sub-solar ($Z$=0.004) metallicity \citep{2005ApJ...622L.117H, 2014ApJ...791...50S} in a locally, metal-poor environment. At the site of this SNR, significant star formation has occurred at metallicities of $Z$=0.004 and $Z$=0.008, with the latter dominating at look back times $\gs$10 Myr. From 100--1000 Myr ago, the SF included a lower metallicity ($Z$=0.001 and $Z$=0.004) component as well as the metal-rich component that contributed at look back times $\ls$500~Myr. Based on the $Z = 0.008$ SFH associated with IKT~6, 46\% of the CC SN progenitors should be stars that have a mass of 8--12.5\,$M_{\odot}$, and 54\% of the massive stars have mass of 12.5--21.5\,$M_{\odot}$, assuming a single star population. This result is consistent with the $\sim$13--15$M_{\odot}$ progenitor suggested by \citet{2005ApJ...622L.117H, 2014ApJ...791...50S}. We note that the SFH of IKT~6 has a strong peak in the 50--200 Myr bin. Assuming a binary star population, this peak could indicate that 8\% of progenitors underwent a delayed CC.

\textbf{IKT~7} (\textit{HFPK~424, SNR~J0051.9$-$7310}): IKT~7 was classified as a SNR by \citet{1983IAUS..101..535I} using X-ray hardness ratios from \emph{Einstein Observatory}. Follow-up {\it XMM-Newton} observations by \citet{haberl12} found that IKT~7 is associated with the 172-s Be/X-ray pulsar AX~J0051.6$-$7311, originally identified with \textit{ASCA} \citep{2000PASJ...52L..37Y}. \citet{2008A&A...485...63F} and \citet{haberl12} did not detect extended emission from the SNR, but the exposures were relatively short ($\sim$10--30~ks) in both studies. As a result the nature of this SNR is uncertain, however, the SNR may be associated with a Be/X-ray pulsar binary system AX~J0051.6$-$7311 \citep{2000PASJ...52L..37Y, haberl12}, indicating the remnant has a CC origin. The region shows strong, metal-rich SF $\sim$30 Myr ago, with weaker but still significant SF 100--500 Myr ago. Within uncertainties, low-metallicity ($Z$=0.004) SF may have occurred $\sim$1 Gyr ago. Based on the SFH, we expect 100\% of the CC SN progenitors to be stars with masses of 8--12.5$M_{\odot}$. As Be/X-ray binaries contain B-star companions of mass 2--16$M_{\odot}$, it is possible that the companion formed in the same population of stars as the progenitor of IKT~7. We note that significant SF occured 50--200 Myr ago, with 7\% of the stellar population undergoing delayed CC assuming a binary stellar population. If the Be/X-ray pulsar system is confirmed to be associated with this remnant, this could suggest that IKT~7 arose from a delayed CC. However, further follow-up of this source is warranted.

\textbf{B0050$-$728} (\textit{DEM~S68SE, HFPK~285, SNR~J0052.6$-$7238, SNR~J005240$-$723820, SMC~258, NS~76}): Detected at both radio \citep{2005MNRAS.364..217F} and optical wavelengths \citep{2007MNRAS.376.1793P}, this very large SNR (diameter$\sim$7\arcmin) shows significant thermal X-ray emission \citep{haberl12} and evidence of [O \textsc{ii}], [S \textsc{ii}], H$\alpha$, and H$\beta$ \citep{2007MNRAS.376.1793P}. \citet{haberl12} suggested that B0050$-$728 is either a large SNR or is one of a pair of SNRs with similar temperatures. This source is located near the H\textsc{ii} region N51 \citep{2014ApJ...795..121L}. The region's SFH exhibits an intense peak of metal-rich SF around $\sim40$ Myr ago and little subsequent SF. The SFH resembles that of the CC SNRs, so it is possible this source arose from a CC event. We estimate that all CC SN progenitors in the region have a mass of 8--12.5\,$M_{\odot}$. 

\textbf{IKT~16} (\textit{HFPK~185/194, SNR~J0058.3$-$7218}): This source was first classified as a SNR using \emph{Einstein} observations \citep{1983IAUS..101..535I} before its shell-type nature was confirmed by radio and H$\alpha$ observations by \citet{1984ApJS...55..189M}. Using \emph{XMM-Newton}, \citet{2004A&A...421.1031V} found  hard X-ray emission at the centre of IKT~16. \citet{2011A&A...530A.132O} found strong evidence of a PWN in both X-ray and radio observations, and they estimated that the NS has a kick velocity of $580\pm100$ km s$^{-1}$. Follow-up \emph{Chandra} observations analysed by \citet{2015A&A...584A..41M} confirmed the presence of a PWN and constrained both the spectral and spin-down properties of the source. \citet{2011A&A...530A.132O} found that the SNR is highly extended, with a radius of 37d$_{60\rm kpc}$ pc, making it one of the largest SNRs in the SMC. \citet{2011A&A...530A.132O} calculated the Sedov age of the SNR to be $\sim15$ kyr. The detection of a PWN associated with the SNR implies that this remnant arises from a CC SN. The thermal X-ray emission from IKT~16 is faint, precluding an estimate of the progenitor mass based on its X-ray properties. Consequently, the SFH in its vicinity yields the first constraints on its progenitor star mass. The region had strong, metal-rich SF 20--100 Myr ago, followed by weak SF across all metallicities. Based on its SFH, the SNR likely arose from a progenitor of mass 8--12.5$M_{\odot}$. IKT~16 also exhibits substantial SF around 50--200 Myr, which could imply that some of the stellar population associated with this remnant underwent delayed CC.

\textbf{IKT~18} (\textit{1E~0057.6$-$7228, HFPK~148, NS~66, SNR~J0059.4$-$7210}): First detected using \emph{Einstein} \citep{1981ApJ...243..736S, 1983IAUS..101..535I}, this SNR is located close to the he bright, young ($\sim$ 3 Myr old) star cluster NGC~346 \citep{1998A&AS..127..119F} that powers the H\textsc{ii} region N66 \citep{1989AJ.....98.1305M, 2014ApJ...795..121L}, and was confirmed as a SNR by \citet{2005MNRAS.364..217F} using ACTA. Coincident with IKT~18 is the multi-star system HD~5980 \citep[e.g.,][]{1997A&A...328..269S}, but the latter source is likely located behind the SNR \citep{2002ApJ...580..225N}. \citet{2004A&A...421.1031V} analysed an \emph{XMM-Newton} observation of the SNR and found that it has a diffuse, centre-filled morphology and ISM-like abundances. \citet{2004A&A...421.1031V} estimated a SNR age of 11 kyr. Due its location close to H\textsc{ii} region NGC~346, \citet{2002PASJ...54...53Y} suggested this SNR arose from a CC SN, while \citet{2007AJ....133...44S} showed that NGC~346 has evidence of significant SF between 3--5 Gyr ago. Based on IKT~18's SFH, we find bursts of SF $\sim$40~Myr ago as well as a sharp peak $\ls$5 Myr ago that likely is responsible for the birth of NGC~346. Based on our estimates, we expect that 91\% of progenitor stars have masses $>$21.5$M_{\odot}$ and 9\% have masses of 8--12.5\,$M_{\odot}$.

\textbf{DEM~S108} (\textit{B0058$-$71.8, HFPK~45, SNR~J0100.3$-$7134}): Classified as a SNR by \citet{1984ApJS...55..189M} and \citet{1982MNRAS.200.1007M} based on radio and optical observations, this faint SNR was first detected in X-rays using \emph{ROSAT} \citep{2000A&AS..142...41H}. Deeper, follow-up observations by \citet{2008A&A...485...63F} using \emph{XMM-Newton} found that its X-ray emission is described by a single temperature, highly absorbed, non-equilibrium ionisation plasma. Due to the low X-ray surface brightness of the SNR, \citet{2008A&A...485...63F} were unable to constrain the elemental abundances. \citet{2008A&A...485...63F} found a well-defined, elliptical shell in 6-cm ATCA radio observations that traces the optical shell found in the Magellanic Cloud Emission Line Survey \citep[e.g.,][]{winkler15}. The stellar cluster Bruck~101 \citep{1976ORROE...1.....B} is located near the southern rim of the SNR and may be where DEM~S108's progenitor star originated. Its SFH exhibits the extensive SF activity in the recent past ($<50$ Myr ago), with an intense burst of metal-rich SF $\sim$8--10 Myr ago. Based on our estimates, 100\% of the CC progenitors arise from stars of mass $>21.5M_{\odot}$. Based on a Salpeter IMF, 34\% of these stars will have a mass $>40M_{\odot}$.

\textbf{IKT~21} (\textit{1E~0101.5$-$7226, HFPK~143, B0101$-$72.4, SNR~J0103.2$-$7209}): This SNR has an incomplete optical and radio shell \citep{1984ApJS...55..189M}, and its faint X-ray emission was first discovered using \emph{Einstein} \citep{1983IAUS..101..535I}. Follow-up \emph{ROSAT} observations by \citet{1994AJ....107.1363H} found that the SNR's X-ray emission is dominated by the Be-pulsar binary system AX~J0103$-$722 \citep{2000ApJ...531L.131I, 2004A&A...414..667H}. Higher resolution \emph{XMM-Newton} observations by \citet{2004A&A...421.1031V} detected faint thermal X-ray emission with a temperature of $kT \sim 0.58$ keV. It is unknown whether AX~J0103$-$722 is associated with the SNR; if so, it would imply that the SNR arose from a CC explosion. The weak X-ray emission of the SNR is associated with enhanced optical ratios that led \citet{1994AJ....107.1363H} to suggest the SNR is expanding into a stellar wind cavity. Similar to IKT~18, IKT~21 is located by the H\textsc{ii} region N66. The elemental abundances of the X-ray plasma have not been constrained, so the progenitor mass is unknown currently, although a CC origin is likely given its proximity to N66 and its possible association with a Be/X-ray binary. Similar to IKT~18, the SFH of IKT~21 is dominated by intense, metal-rich SF $<8$ Myr ago, a burst of SF $\sim$40 Myr ago, and very little SF prior to then. It is expected that 92\% of the progenitor stars have masses $>$21.5$M_{\odot}$ and 8\% have masses of 8--12.5\,$M_{\odot}$.

\textbf{HFPK~334} (\textit{SNR~J0103.5$-$7247}): First detected using {\it ROSAT} \citep{1999A&AS..136...81K}, HFPK~334 is unusual due to the fact that it emits radio and X-rays \citep{2000A&AS..142...41H, 2008A&A...485...63F, 2014AJ....148...99C} but shows no evidence of optical emission \citep{2007MNRAS.376.1793P}. \citet{2008A&A...485...63F} detected non-thermal, point-like emission at the centre of the SNR which they suggested could be a putative PWN. Follow-up observations by \citet{2014AJ....148...99C} confirmed the presence of the point source but that it was a background object not associated with the SNR. Furthermore, they found that the SNR's diffuse X-ray emission is best described by a thermal, non-equilibrium ionization plasma model with SMC ISM abundances.  \citet{2014AJ....148...99C} estimated an age of $\sim1800$ years for SNR, but they cautioned that the age may be an underestimate because the SNR is expanding into a low-density environment.

\textbf{1E 0102.2$-$7219} (\textit{DEM~S124, IKT~22, B0102$-$7219. 1E~0102$-$72.3, HFPK~107, SNR~0102$-$72.3, SNR~J0104.0$-$7202}): First discovered in X-rays using \emph{Einstein} \citep{1981ApJ...243..736S}, 1E 0102.2$-$7219 is the brightest SMC SNR in X-rays. Since its discovery, this SNR has been extensively studied in multiple wavelengths. The SNR is oxygen-rich, based on the detection of filamentary [O \textsc{iii}] emission \citep{1981ApJ...248L.105D,1983ApJ...268L..11T}. \emph{Hubble} Space Telescope observations indicated the presence of O, Ne and Mg, leading \citet{1989ApJ...338..812B, 2000ApJ...537..667B} to suggest that it was a Type Ib asymmetric, bipolar SN of a 25--35 $M_{\odot}$ Wolf-Rayet star \citep{2010ApJ...721..597V}. Follow-up observations using \emph{Spitzer} \citep{2005ApJ...632L.103S, 2009ApJ...700..579R}, \emph{FUSE} \citep{2006ApJ...642..260S}, \emph{XMM-Newton} \citep{2006ApJ...642..260S}, \textit{Chandra } \citep{2000ApJ...534L..47G, 2001A&A...365L.231R, 2001AIPC..565..230D, 2004ApJ...605..230F, 2017A&A...597A..35P} confirmed the presence of O, Ne and Mg in the ejecta and found little-to-no emission from Fe or other heavy elements. Deep \emph{Chandra} observations have not detected a neutron star \citep{2010ApJ...715..908R}, however recently \citep{2018arXiv180301006V} reported the detection of a possible compact central object associated with the remnant that has similar properties to the central compact object of Cas A. Using proper motion measurements of the X-ray filaments of 1E~0102.2$-$7219, \citet{2000ApJ...543L..61H} inferred a SNR age of $\sim$1000 years, while \citet{2006ApJ...641..919F} interred an age of $\sim$2050 years from optical filaments. 1E~0102.2$-$7219 is in an active star-forming area of the SMC. The SFH is dominated by extensive, metal-rich ($Z$=0.008) SF for lookback times $>5$ Myr. Although uncertain, it is possible that lower metallicity ($Z$=0.004) SF also occurred. Based on Figure \ref{SFH1}, the SFH of the vicinity of 1E~0102.2$-$7219 suggests that 31\% of the CC progenitors have a mass of 8--12.5\,$M_{\odot}$, 20\% have a mass of 12.5--21.5\,$M_{\odot}$, and 49\% have a mass $>21.5M_{\odot}$, using a single star population. Assuming a Salpeter IMF, 69\% of the $>21.5M_{\odot}$ stars would have a mass of 21.5--40$M_{\odot}$, so it is possible that the progenitor was a 25--35$M_{\odot}$, as suggested by \cite{2000ApJ...537..667B}. We also note that significant SF occurred 50--200 Myr ago in the subcell of this source. Assuming a binary stellar population DTD, it is possible that 6\% of the stellar population underwent a delayed CC.

\textbf{IKT~23} (\textit{1E~0103.3$-$7240, DEM~S125, HFPK~217, SNR~0103$-$72.6, SNR~J0105.1$-$7223}): IKT~23 is the second brightest X-ray SNR in the SMC \citep{1981ApJ...243..736S, 1987ApJ...317..152B, 1997PASP..109...21C,  2000A&AS..142...41H, 2004A&A...421.1031V}. It shows a well-defined shell morphology in X-rays \citep{2002PASJ...54...53Y, 2003ApJ...598L..95P, 2004A&A...421.1031V}, and the SNR is faint in the radio \citep{1982MNRAS.200.1007M} and optical \citep{1984ApJS...55..189M}. This SNR is located close to the H\textsc{ii} region DEM~S125, and X-ray observations with \emph{ASCA} \citep{2002PASJ...54...53Y}, \emph{Chandra} \citep{2003ApJ...598L..95P} and \emph{XMM-Newton} \citep{2004A&A...421.1031V} showed the SNR has enhanced abundances of O and Ne and marginally enhanced heavier elements compared to SMC ISM abundances. Thus, authors have noted the similarity of IKT~23 with both 1E 0102.2$-$7219 and the Galactic SNR G292.0+1.8 \citep{2002ApJ...564L..39P}.  By comparing the elemental abundances of this remnant with Type II SN nucleosynthesis models, \citet{2003ApJ...598L..95P} suggested that the SNR resulted from a CC of a $>$18 $M_{\odot}$ progenitor. Similar to 1E~0102.2$-$7219, the SFH of IKT~23 has significant metal-rich SF at lookback times $\gs$10~Myr. Within uncertainties, there may have been $Z$=0.004 SF as well, and this SF dominated for earlier lookback times of 1--5~Gyr. The SFR $\ls$50 Myr ago implies that 100\% of the CC SN progenitors in this subcell have masses of 8--12$M_{\odot}$, below the $\sim$18$M_{\odot}$ suggested by \citet{2003ApJ...598L..95P}. The vicinity of IKT~23 exhibits substantial SF $\gs$50 Myr ago, which from our analysis suggests that a significant fraction of the progenitors underwent delayed CC.

\textbf{DEM~S128} (\textit{SNR 0103$-$72.4, B0104$-$72.2, ITK~24, HFPK~145, SNR~J0105.4$-$7209}): First classified as a SNR associated with the H\textsc{ii} region of the same \citep{1983IAUS..101..535I}, this source has been observed many times in X-rays \citep{1987ApJ...317..152B, 1992ApJS...78..391W, 2000A&A...353..129F, 2002PASJ...54...53Y, 2004A&A...421.1031V}. Follow-up observations in radio \citep{1997A&AS..121..321F, 2000A&A...353..129F, 2005MNRAS.364..217F}, optical \citep{2000A&A...353..129F, 2007MNRAS.376.1793P} and infrared \citep{1989A&AS...79...79S} confirmed the SNR classification of the object. Located near the SNR is a Be/X-ray binary AX~J0105$-$722 \citep[see e.g.,][and reference therewithin]{2016A&A...586A..81H}. \citet{2002PASJ...54...53Y} suggested the X-ray binary is associated with SNR DEM~S128, but other studies indicate it may not be tied to the SNR \citep{2000A&A...353..129F,2004A&A...421.1031V}. The X-ray spectrum of DEM~S128 from {\it XMM-Newton} observations showed an Fe abundance three times that of the SMC ISM \citep{2004A&A...421.1031V}. This Fe enhancement was confirmed by \citet{2015ApJ...803..106R} using \emph{Chandra} observations. Due to its centre-filled X-ray morphology, large optical and radio diameter \citep{1984ApJS...55..189M, 2000A&A...353..129F}, and enhanced Fe abundance, \citet{2004A&A...421.1031V} and \citet{2015ApJ...803..106R} suggested the SNR is an old remnant from a Type Ia SN. Details of its SFH is found in Section \ref{iacc}.

\textbf{DEM~S130}, (\textit{SNR~J0105.6$-$7204}): Detected as a shell-type SNR in radio using ACTA and in optical \citep{2005MNRAS.364..217F}, this source has not been detected in X-rays \citep{2000A&AS..142...41H,haberl12} down to a surface-brightness limit of $\sim10^{-14}$~erg~cm$^{-2}$~s$^{-1}$~arcmin$^{-2}$. The SNR is a part of the H\textsc{ii} region, ATCA SMC~444 (J010539$-$720341) \citep{2005MNRAS.364..217F} and is associated with the H\textsc{ii} region of the same name \citep{2012ApJ...755...40P}. It is also located near the SNRs IKT~25 and DEM~S128. The region's SFH shows an intense burst of SF $<$50 Myr ago. Thus, DEM~S130 is likely from a CC SN. 76\% of the CC progenitors coincident with DEM~S130 are stars of mass $>21.5M_{\odot}$, while 15\% arise from 12.5--21.5\,$M_{\odot}$ and 9\% arise from 8--21.5\,$M_{\odot}$. Of the $>21.5M_{\odot}$ progenitors, 34\% of these stars will have a mass $>40M_{\odot}$ assuming a Salpeter IMF.

\textbf{IKT~25}, (\textit{B0104$-$72.3, HFPK~125, SNR~0104$-$72.3, SNR~J0106.2$-$7205}): The fourth brightest X-ray SNR in the SMC, IKT~25's explosive origin has been debated in the literature. It was discovered in optical wavelengths by \citet{1984ApJS...55..189M}, and it has been studied extensively in X-rays using \emph{ROSAT} \citep{1994AJ....107.1363H}, \emph{XMM-Newton} \citep{2004A&A...421.1031V}, \emph{Chandra} \citep{2011ApJ...731L...8L, 2014ApJ...788....5L, 2015ApJ...803..106R}, and \emph{Suzaku} \citep{2016PASJ...68S...9T}. IKT~25 has also been observed in the infrared \citep{2007PASJ...59S.455K}, follow-up optical observations \citep{1984ApJS...55..189M, 1994AJ....107.1363H, 2007MNRAS.376.1793P}, and in the radio \citep{ 2005MNRAS.364..217F}. The SNR was suggested to arise from a Type Ia SN based on its Balmer-dominated optical spectrum, and \citet{2011ApJ...731L...8L} argued the explosion was a prompt Type Ia explosion in a star-forming region \citep{2007PASJ...59S.455K}. Based on enhanced Fe abundances in X-ray observations, \citet{2004A&A...421.1031V} and \citet{2015ApJ...803..106R} also concluded IKT~25 had a Type Ia origin. However, based on its association with a star-forming region, the detection of a bright IR shell tracing out H$\alpha$ emission, its asymmetric morphology, its abundances of intermediate-mass elements, other works have claimed it is more consistent with a CC origin \citep{1994AJ....107.1363H, 2007PASJ...59S.455K, 2014ApJ...788....5L,2016PASJ...68S...9T}. \citet{2014ApJ...788....5L} noted that \citet{2011ApJ...731L...8L} adopted the incorrect SMC ISM abundances in their X-ray spectral fits, leading to the mistyping of this source as a Type Ia SNR. Based on their abundance estimates, the SNR's elongated morphology, and the star-formation history at the site of IKT~25, \citet{2014ApJ...788....5L} suggested that the SNR arose from a bipolar Type Ib/Ic CC SN of a $\sim$25 $M_{\odot}$ progenitor. We find that the SFH associated with IKT~25\footnote{We note that \cite{2014ApJ...788....5L} also characterised the SFH of IKT~25, but they selected the wrong subcell (an adjacent one) for this calculation due to a coordinate error. Nonetheless, their conclusion that the SNR likely arose from a $\sim$25$M_{\odot}$ progenitor still holds.} is dominated by intense, metal-rich SF $\sim$8~Myr ago that peaked after a steadily increasing SFR since $\sim$90 Myr ago. As a result, 76\% of the CC SN progenitors associated with IKT~25 are expected to arise from stars of mass $>21.5M_{\odot}$. Assuming a Salpeter IMF, 69\% of the $>21.5M_{\odot}$ stars would have a mass between 21.5--40\,$M_{\odot}$.

\textbf{N83C}, (\textit{NS~83, SNR~J0114.0$-$7317, DEM~S147, SNR~J011333$-$731704, B0113$-$729, SMC~B0112$-$7333, NGC~456, Nail~148, SMC~547, NS~83(A,C)}): Very little is known about this SNR. This source was suggested by \citet{2005MNRAS.364..217F} to be a shell-type SNR candidate (ACTA~SMC~547) that is possibly interacting with a nearby molecular cloud. \citet{haberl12} detected no X-rays from the position of the source using \emph{XMM-Newton}, and \citet{2015ApJ...799..158T} showed that the SNR has a high dust-to-gas ratio, indicating that it is in a high-density environment. N83C is associated with the H\textsc{ii} region DEM~S147 \citep{1956ApJS....2..315H,2012ApJ...755...40P} and is located within a bright CO complex in the SMC Wing that shows evidence of active star formation \citep[e.g.,][]{2003ApJ...595..167B}. Due to its location close to an active star-forming region and that it is found in a high-density environment \citep{2003ApJ...595..167B, 2015ApJ...799..158T}, it is possible this remnant arises from a CC SN. Its SFH supports this conclusion, with a peak in the SFR $\sim$9 Myr ago and limited SF otherwise. We expect that 100\% of the CC SN progenitors in this sub-cell have a mass $>21.5M_{\odot}$. Adopting a Salpeter IMF, 34\% of these stars will have a mass $>40M_{\odot}$.

\section{Candidate SNRs}
\begin{table}[t!]
	\begin{center}
		\caption{Candidate SNRs also suggested in the literature. \label{othersnrs}}
		\begin{tabular}{lccc}
			\toprule
			Name & R.A. & Dec. & Size (arcmin)\\
			\midrule
			SXP~1062 SNR& $01^{h} 29^{m} 12^{s}$&$-73^{d} 32^{m} 02^{s}$ & 14\\
			XTE~J0111.2$-$7317 &$01^{h} 11^{m} 09^{s}$&$-73^{d} 16^{m} 46^{s}$ & 0.15\\
			NS~66D & $00^{h} 57^{m} 46^{s}$&$-72^{d} 14^{m} 04^{s}$ & 4.8\\
			DEM~S25 & $00^{h} 41^{m} 00^{s}$&$-73^{d} 36^{m} 48^{s}$ & 1.7\\
			DEM~S135 & $01^{h} 08^{m} 20^{s}$&$-71^{d} 59^{m} 57^{s}$ & 3.4\\
			XMMU J0049.0$-$7306 & $00^{h} 49^{m} 00^{s}$&$-73^{d} 06^{m} 17^{s}$ & 1.5\\
			XMMU J0056.5$-$7208 & $00^{h} 56^{m} 30^{s}$&$-72^{d} 08^{m} 12^{s}$ & 3.4\\
			\bottomrule
		\end{tabular}
	\end{center}
\end{table}

For reference, we list below candidate SMC SNRs that have been suggested in the literature. We do not include these sources in our analysis as more detailed follow-up observations are required to confirm their SNR nature.

\textbf{SXP~1062~SNR:} SXP~1062 is a Be/X-ray binary that is one of three SMC X-ray pulsars with a pulse period greater than 1000~s. Its optical counterpart is a B0-0.5(III)e$+$ star \citep{2012MNRAS.420L..13H}. \citet{2012MNRAS.420L..13H} discovered faint optical emission surrounding the binary which they suggested was from a SNR. Using multi-wavelength follow-up, \citet{2012A&A...537L...1H} confirmed the presence of a shell-type SNR that emits at radio, optical and X-ray wavelengths. SXP~1062 is found close to the projected centre of this candidate SNR, and \citet{2012MNRAS.420L..13H,2012A&A...537L...1H} suggested that the age of the SNR and pulsar are both $<25$ kyr. Due to the faintness of the thermal X-ray emission, \citet{2012A&A...537L...1H} were unable to constrain the presence of ejecta from this source. This source is near to a star-forming region NGC~602.

\textbf{XTE J0111.2-7317}: Originally discovered as an X-ray pulsar with a 31 s period using RXTE \citep{1998IAUC.7048....1C, 1998IAUC.7062....1C}, \citet{2000MNRAS.314..290C} followed up this source using both optical and IR observations and confirmed this object was in a Be/X-ray binary with a main sequence B0-B2 star. From their optical observations, \citet{2000MNRAS.314..290C} also found evidence of H$\alpha$ emission which they suggest is either a SNR, pulsar wind bow shock nebula of H\textsc{ii} region. Further follow-up observations seem to suggest that the nebulosity surrounding XTE J0111.2-7317 is likely just a H\textsc{ii} region \citep{2003MNRAS.344.1075C}.

\textbf{NS~66D} \textit{(B0056$-$724, SNR~J005800$-$721101, XMMU~J0057.7$-$7213, ATCA~345)}: Located in the southwest bar of the SMC, this source was suggested to be a large (114\arcsec) SNR by \citet{2005MNRAS.364..217F} and \citet{2007MNRAS.376.1793P} based on the the radio spectral index derived from ATCA and the detection of radio shell and optical emission. No follow-up observations have been taken to confirm the nature of this source, and it may be associated with the X-ray SNR candidate XMMU~J0057.7$-$7213 suggested by \citet{haberl12}.

\textbf{DEM~S25}, \textit{DEM~S142, N17, SNR~J004640$-$733150, IJL~J004643$-$733112, NGC~265}:  Similar to NS~66D, this source was suggested be a SNR based its radio spectral index derived from ATCA \citet{2005MNRAS.364..217F}. However, this source has not been followed up to confirm its SNR origin.

\textbf{DEM~S135}, \textit{N 80 (80A), S 23, SMC~B0106$-$7215, SNR~J010819$-$715956}: As DEM~S25 and SN~66D, this source was classified as a SNR candidate by \citet{2005MNRAS.364..217F}.

\textbf{XMMU~J0049.0$-$7306:} Discovered in their X-ray study of the SMC, \citet{haberl12} suggested this source was a SNR, but no follow-up observations have been taken to confirm the nature of this source.

\textbf{XMMU~J0056.5$-$7208:} \citet{haberl12} discovered this source and suggested it was a SNR based on detection of thermal X-ray emission. The low-count statistics of the {\it XMM-Newton} observation precluded constraints on the hot plasma properties of the source. Using MCELS, they did confirm that an elliptical shell in H$\alpha$ and [S \textsc{ii}] is detected. Follow-up observations are needed to confirm the nature of this source.

\end{appendix}

\bibliography{reference.bib}
\bibliographystyle{apj}

\end{document}